\documentclass[aps,prb,twocolumn,showpacs,superscriptaddress,amsmath,amssymb]{revtex4-1}

\usepackage{graphicx}
\usepackage[colorlinks=true,linkcolor=blue,citecolor=blue,urlcolor=blue]{hyperref}

\def \b {\beta}
\def \d {\delta}
\def \D {\Delta}

\def \ve {\varepsilon}

\def \g {\gamma}

\def \l {\lambda}
\def \L {\Lambda}
\def \o {\omega}
\def \O {\Omega}
\def \t {\theta}

\def \z {\zeta}
\def \dag {\dagger}

\def \p {\partial}
\newcommand{\pd}[2]{\fr{\p {#1}}{\p {#2}}}

\def \del {\nabla}

\def \apx {\approx}

\def \til {\tilde}

\def \dag {\dagger}

\newcommand{\intvrdp}{\int_{\mbf r\mbf r'}}

\def \inttauvr {\int_{\tau\mbf r}}

\def \rar {\rightarrow}

\def \la {\langle}
\def \ra {\rangle}
\def \fr {\frac}
\def \lf {\left}
\def \ri {\right}

\newcommand{\ket}[1]{|#1\ra}
\newcommand{\braket}[3]{\la#1|#2|#3\ra}

\def \Tr {\operatorname{Tr}}
\def \bece {\begin{center}}
\def \ence {\end{center}}
\def \beeq {\begin{equation}}
\def \eneq {\end{equation}}
\def \beal {\begin{aligned}}
\def \enal {\end{aligned}}
\def \bega {\begin{gathered}}
\def \enga {\end{gathered}}
\def \benu {\begin{enumerate}}
\def \ennu {\end{enumerate}}
\def \beit {\begin{itemize}}
\def \enit {\end{itemize}}
\def \bede {\begin{description}}
\def \ende {\end{description}}
\def \betb {\begin{tabular}}
\def \entb {\end{tabular}}
\def \bear {\begin{array}}
\def \enar {\end{array}}

\def \mbf {\mathbf}
\def \mbb {\mathbb}
\def \mrm {\mathrm}

\def \mca {\mathcal}

\def \bsb{\boldsymbol}

\begin{document}


\title{Exotic Superconductivity with Enhanced Energy Scales in Three Band Crossing Materials}

\author{Yu-Ping Lin}
\affiliation{Department of Physics, University of Colorado, Boulder, CO, 80309}
\author{Rahul M. Nandkishore}
\affiliation{Department of Physics, University of Colorado, Boulder, CO, 80309}
\affiliation{Center for Theory of Quantum Matter, University of Colorado, Boulder, CO, 80309}

\date{\today}

\begin{abstract}
Three band crossings can arise in three dimensional quantum materials with certain space group symmetries. The low energy Hamiltonian supports spin {\it one} fermions and a flat band. We study the pairing problem in this setting. We write down a minimal BCS Hamiltonian and decompose it into spin-orbit coupled irreducible pairing channels. We then solve the resulting gap equations in channels with zero total angular momentum. We find that in the spin singlet channel (and also in an unusual `spin quintet' channel), superconductivity is enormously enhanced, with a possibility for the critical temperature to be {\it linear} in interaction strength. Meanwhile, in the spin triplet channel, the superconductivity exhibits features of conventional BCS theory due to the absence of flat band pairing. Three band crossings thus represent an exciting new platform for realizing exotic superconducting states with enhanced energy scales. We also discuss the effects of doping, nonzero temperature, and of retaining additional terms in the $\mathbf{k} \cdot \mathbf{p}$ expansion of the Hamiltonian.  
\end{abstract}

\maketitle

\section{Introduction}

The emergence of topological band structures has provided a new playground for many body theory (see e.g. Ref.~\onlinecite{Armitege2018RMP} and references contained therein). While most effort has focused on systems with two band crossings (topological insulator surface states, Weyl semimetals) or four band crossings (graphene, Dirac semimetals), band crossings of {\it odd} numbers of bands can also exist, stabilized by certain space group symmetries \cite{Bradlyn2016Sci}. A concrete example of these novel materials is provided by the time reversal (TR) symmetric materials with space group symmetry 199, where three band crossings arise. 

It is by now well known that topological band structures present intriguing platforms for realizing novel correlated states of matter. For example, even restricting ourselves purely to superconductors, superconducting topological insulator surface states support Majorana fermions in vortex cores \cite{FuKane}, superconducting Weyl semimetals realize `monopole harmonic pairing' \cite{Li2018PRL}, superconductivity in certain four band crossing materials realizes unusual `high spin' superconducting states \cite{congjun, agterberg, paglione, Herbut, Savary2017PRB, Boettcher, Venderbos, Ghorashi2017PRB, Roy2017Arxiv}, and nodal line semimetals support unusual chiral states with topological surface superconductivity \cite{Nandkishore, SurNandkishore, WangNandkishore, Wangetal}. However, investigations have thus far been restricted to two and four band crossings, and the three band crossing setting is one in which correlated states in general, and the pairing problem in particular, has scarcely been explored. 

Here we initiate the exploration of many body theory in three band crossing systems by studying {\it superconductivity} in this setting. The BCS problem in this setting is interesting for at least two distinct reasons. Firstly, the effective theory of the three band crossing (as encapsulated in the $\mbf{k} \cdot \mbf{p}$ Hamiltonian) is a theory of {\it spin one} fermions \cite{Bradlyn2016Sci}. The pairing problem for spin one fermions has not previously been studied, as far as we are aware. Additionally, the $\mbf{k} \cdot \mbf{p}$ Hamiltonian for three band crossings supports a {\it flat} band with a concomitant large density of states. This feature can bring about dramatical enhancement of superconductivity \cite{volovik, Uchoa2013PRL}. 

In this paper, we focus on an idealized model (the leading order $\mbf{k} \cdot \mbf{p}$ Hamiltonian from Ref.~\onlinecite{Bradlyn2016Sci}), which has full spherical symmetry. This idealized problem is sufficient to capture the key features of the pairing problem in this setting. In the presence of full spherical symmetry (but with significant spin-orbit coupling), the pairing states are classified by quantum numbers $(L,S,J, M_J)$, where $L$ is the orbital angular momentum, $S$ the total spin, $J$ the total angular momentum, and $M_J$ the projection of the total angular momentum along the (arbitrarily chosen) $\hat z$-axis. Unusually, because we are dealing with spin one fermions, there is a possibility of a spin quintet ($S=2$) pairing state. We write down a general action, including short range attractive interaction, and decompose it into different irreducible pairing channels labelled by $(L,S,J)$, adapting the analysis from Ref.~\onlinecite{Savary2017PRB}. We then solve the (weak coupling) pairing problem in channels with $J=0$, such that the solutions are nondegenerate. The restriction to channels with $J=0$ is purely for simplicity, and is sufficient to highlight the key features. We investigate the superconductivity when the Fermi level lies exactly at the band crossing points, and discuss the effects of finite temperature and finite displacement of Fermi level from band crossing points. Interesting results show up when the Fermi level is at the band crossing points. Otherwise one may simply project the weak coupling pairing problem onto a single linearly dispersing band, and the novel features of the spin one problem are eliminated. In particular, the $s$-wave spin singlet $(0,0,0)$ and $d$-wave spin quintet $(2,2,0)$ pairing states display an enormous enhancement of the superconductivity. When Fermi level lies at band crossing points, the critical temperature is {\it linear} in interaction strength, instead of being exponentially small as in the standard BCS problem. This is reminiscent of superconductivity in flat bands \cite{volovik, Uchoa2013PRL}. The linear scaling still shows up when the Fermi level is slightly shifted once the interaction is strong enough. Meanwhile, the pairing problem in the $p$-wave spin triplet $(1,1,0)$ pairing channel manifests conventional BCS superconductivity due to the absence of flat band pairing. In addition to the perfect flat band setup, the effect of finite band curvature on superconducting behavior is also examined. The modifications of weak coupling theories in $s$-wave spin singlet and $d$-wave spin quintet pairing states are discovered. We conclude with a discussion of our results and of future directions.

\section{Setup}
\label{SecPS1SM}
\subsection{Noninteracting Hamiltonian} 

\begin{figure}[t]
\includegraphics[scale=0.5]{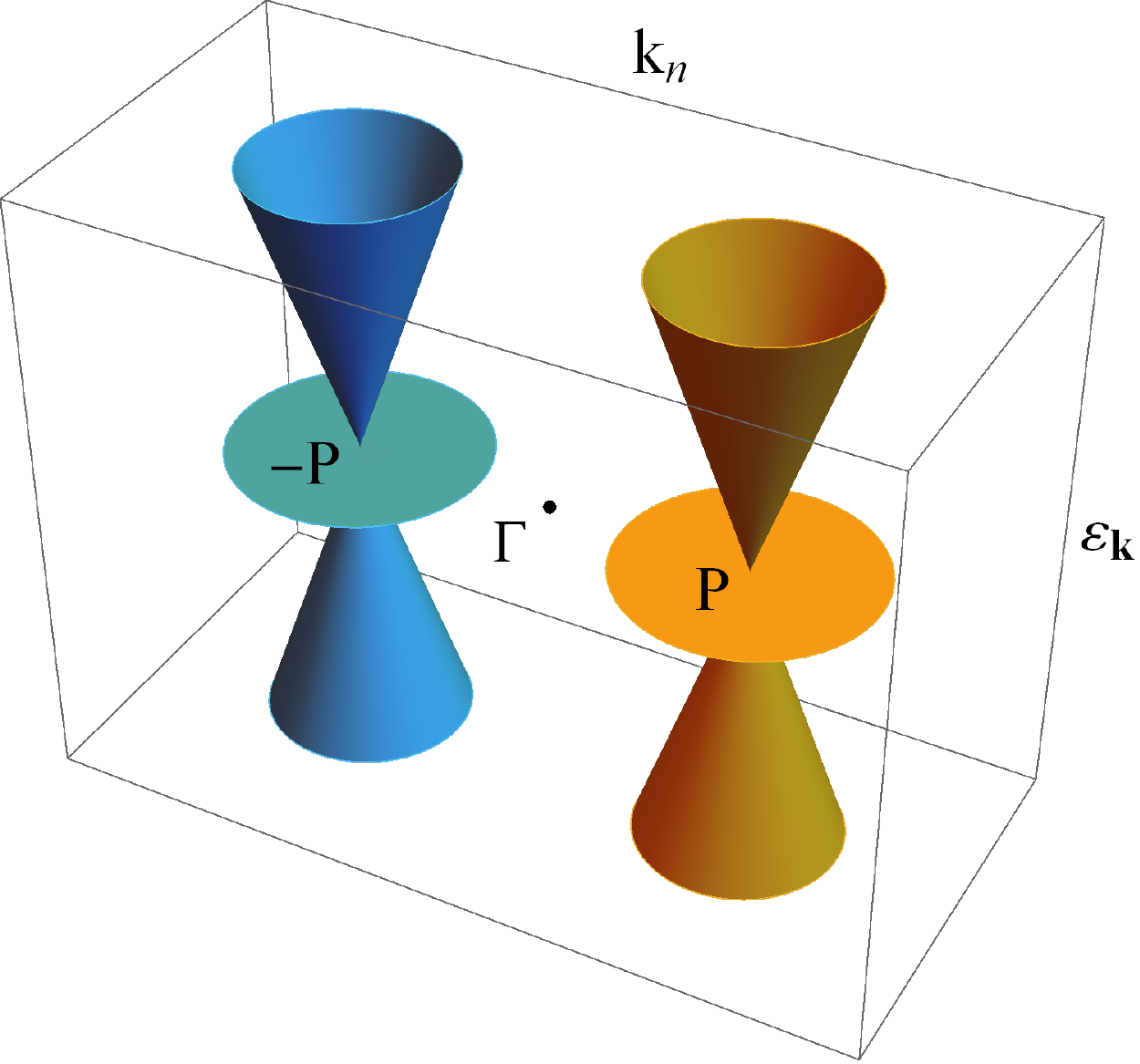}
\caption{\label{Fig3BTP} Schematic illustration of a pair of three band crossing points $\pm P: \mbf k=\pm\mbf k_0$ in the bulk Brillouin zone. The axis $k_n$ is in the direction $\hat k_0$. The flat bands at the three band crossing points $\pm P$ can produce nontrivial superconductivities with the pairing between states near $\pm P$.}
\end{figure}

In a TR symmetric material which exhibits space group symmetry 199, there can exist a pair of three band crossing points at the TR noninvariant $P$-point and its TR partner $-P$ in the bulk Brillouin zone (Fig.~\ref{Fig3BTP}). A minimal model $\mbf{k} \cdot \mbf{p}$ Hamiltonian about the $P$-point $\mbf k=\mbf k_0$ is \cite{Bradlyn2016Sci}
\beeq
\label{EqHP}
H_P(\mbf k) = v\lf(\mbf k-\mbf k_0\ri)\cdot\mbf S,
\eneq
where $S^i$'s are the spin-$1$ operators
\beeq
\bega
S^x = \fr{1}{\sqrt{2}}\lf(\bear{ccc}0&1&0\\1&0&1\\0&1&0\enar\ri),\quad
S^y = \fr{1}{\sqrt{2}}\lf(\bear{ccc}0&-i&0\\i&0&-i\\0&i&0\enar\ri),
\\
S^z = \lf(\bear{ccc}1&0&0\\0&0&0\\0&0&-1\enar\ri),
\enga
\eneq
and $v$ denotes the effective velocity. Notice the choice $\hbar=1$ in the setup. The Hamiltonian Eq.~(\ref{EqHP}) exhibits two linearly dispersing bands and a flat band
\beeq
\ve_{\mbf k}^\pm = \pm v\lf|\mbf k-\mbf k_0\ri|,\quad
\ve_{\mbf k}^0 = 0.
\eneq
Such three band crossing structure exhibits nontrivial topological features. The topological property of each band can be captured by the Berry monopole charge $C^a=(1/2\pi)\oint d\mbf S_{\mbf k}\cdot\bsb\O_{\mbf k}^a$ at the band crossing point $P$, where the integral goes over an arbitrary closed surface enclosing $P$. The Berry flux $\bsb\O_{\mbf k}^a=\del_{\mbf k}\times\mbf A_{\mbf k}^a$ is calculated from the Berry connection $\mbf A_{\mbf k}^a=\braket{u_{\mbf k}^a}{i\del_{\mbf k}}{u_{\mbf k}^a}$ where $\ket{u_{\mbf k}^a}$ denotes the state on the $a$-th band. One can check that the two linearly dispersing bands carry Berry monopole charge $C^\pm=\mp2$, implying nontrivial topological structure, while the flat band is topologically trivial \cite{Bradlyn2016Sci}.

The operator $\mca T$ of TR transformation is defined as $\mca T = \g K$, where $K$ is the complex conjugate operator and $\g$ is an unitary operator
\beeq
\g
= i\exp\lf(i\pi S^y\ri)
= \lf(\bear{ccc}0&0&i\\0&-i&0\\i&0&0\enar\ri).
\eneq
The additional factor $i$ is introduced so as to fulfill the condition $\mca T^2=-1$ for spin-orbit coupled electrons. Under TR transformation, the spin operators are transformed as
\beeq
\label{EqTRST}
\mbf S\rar\mca T\mbf S\mca T^{-1} = \g\mbf S^*\g^\dag = -\mbf S.
\eneq
To fulfill the TR symmetry, the Hamiltonian $H_{-P}(\mbf k)$ at $-P$ has to satisfy
\beeq
\label{eq:TRS}
\mca TH_P^*(\mbf k)\mca T^{-1} = H_{-P}(-\mbf k),
\eneq
which implies
\beeq
\label{EqHNP}
H_{-P}(-\mbf k) = v\lf[-\mbf k-\lf(-\mbf k_0\ri)\ri]\cdot\mbf S.
\eneq
The Hamiltonians $H_{\pm P}(\mbf k)$ take the same form in terms of $\d\mbf k=\mbf k-(\pm\mbf k_0)$. The Berry monopole charges at $\pm P$ are the same, which is a general feature of TR symmetric systems.

\subsection{Noninteracting Action}
\label{SecAc}
It is convenient to recast the problem in (Matsubara) coherent path integral language \cite{AltlandSimons, Coleman}, with noninteracting action
\beeq
\label{EqNIAO}
S_0 = \sum_{a=\pm}\inttauvr\psi_a^\dag\lf[\p_\tau+v(-i\del)\cdot\mbf S-\mu\ri]\psi_a.
\eneq
The $3$-component vectors of Grassmann numbers $\psi_\pm$'s characterize the states near $\pm P$ with a momentum difference $\pm\mbf k$, respectively. The chemical potential is denoted by $\mu$. For the description at temperature $T$, the periodic domain in temporal direction is defined by the inverse temperature $\b=1/k_BT$, where $k_B=1$ is chosen. 

The BCS superconducting states are generated by the Cooper pairing between $\psi_+^\dag$ near $P$ and $\g(\psi_-^\dag)^T$ near $-P$. For future convenience, we rewrite the $(\psi_-)$-part of the noninteracting action Eq.~(\ref{EqNIAO}) in terms of $\g(\psi_-^\dag)^T$.
The noninteracting action Eq.~(\ref{EqNIAO}) may then be expressed as \cite{AltlandSimons,Coleman}
\beeq
\label{EqNIANS}
S_0
=-\inttauvr
\Psi^\dag(\mbf r,\tau)\mca G_0^{-1}(\mbf r,\tau)\Psi(\mbf r,\tau)
,
\eneq
where we have introduced the Nambu spinor
\beeq
\Psi(\mbf r,\tau) = \lf(\bear{c}\psi_+(\mbf r,\tau) \\ \g[\psi_-^\dag(\mbf r,\tau)]^T\enar\ri),
\eneq
and $\mca G_0^{-1}$ is the noninteracting inverse Gor'kov Green's function
\beeq
\mca{G}_0^{-1}
=
\lf(\bear{cc}
-\p_\tau-v(-i\del)\cdot\mbf S+\mu & 0 \\
0 & -\p_\tau-v(i\del)\cdot\mbf S-\mu
\enar\ri)
.
\eneq

It is now convenient to proceed to a momentum space representation. Consider the Fourier transform of the Nambu spinor
\begin{align}
\label{EqFTkt}
\Psi(\mbf r,\tau) = \fr{1}{\sqrt{\mca V}}\sum_{\mbf k}\Psi_{\mbf k}(\tau)e^{i\mbf k\cdot\mbf r},
\\
\Psi_{\mbf k}(\tau) = \lf(\bear{c}\psi_{+,\mbf k}(\tau)\\\g[\psi_{-,-\mbf k}^\dag(\tau)]^T\enar\ri),
\end{align}
where $\mca V$ is the spatial volume. The noninteracting action Eq.~(\ref{EqNIANS}) becomes
\beeq
\label{EqNIANSMS}
S_0 = -\int_\tau\sum_{\mbf k}\Psi_{\mbf k}^\dag(\tau)\mca G_0^{-1}(\mbf k,\tau)\Psi_{\mbf k}(\tau),
\eneq
with the inverse Gor'kov Green's function
\beeq
\mca{G}_0^{-1}
=
\lf(\bear{cc}
-\p_\tau-v\mbf k\cdot\mbf S+\mu & 0 \\
0 & -\p_\tau+v\mbf k\cdot\mbf S-\mu
\enar\ri)
.
\eneq

\subsection{Interactions and Irreducible Pairing Channels}
\label{sec:intpairchan}

To complete the setup of the pairing problem, a short range attractive density-density interaction is now introduced to couple the states near $\pm P$
\beeq
\label{EqIntAc}
S_\mrm{int}
= -\fr{1}{2}\int_\tau\intvrdp V(\mbf r-\mbf r')\psi_+^\dag(\mbf r)\psi_+(\mbf r)\psi_-^\dag(\mbf r')\psi_-(\mbf r').
\eneq
The minus sign indicates an attractive interaction.

We decompose the density-density term into a summation over different irreducible spin representations, which can be regarded as a kind of Fierz identity \cite{Savary2017PRB}. After projection on the Cooper channel, the four fermion interaction in this setting takes the form
\begin{align}
&\psi_{+,\mbf k}^\dag\psi_{+,\mbf k'}\psi_{-,-\mbf k}^\dag\psi_{-,-\mbf k'}
\\
&= \psi_{+,\mbf k}^\dag\psi_{+,\mbf k'}\psi_{-,-\mbf k}^\dag\g^T\g^*\psi_{-,-\mbf k'}
\\
&= \psi_{+,\mbf k}^\dag E_{mn}[\g(\psi_{-,-\mbf k}^\dag)^T][\g(\psi_{-,-\mbf k'}^\dag)^T]^\dag E_{nm}\psi_{+,\mbf k'}
.
\end{align}
In the last line, each matrix $E_{mn}$ has only one nonzero element $1$ as the $(m,n)$-th entry. The summation over $E_{mn}$'s is equivalent to a summation over the $(2S+1)$-component vector $\vec{\mca M}_S$'s
\beeq
\beal
&\psi_{+,\mbf k}^\dag\psi_{+,\mbf k'}\psi_{-,-\mbf k}^\dag\psi_{-,-\mbf k'}
\\
&= \fr{1}{3}\sum_S
\psi_{+,\mbf k}^\dag\vec{\mca M}_S[\g(\psi_{-,-\mbf k}^\dag)^T]
\cdot
[\g(\psi_{-,-\mbf k'}^\dag)^T]^\dag\vec{\mca M}_S^\dag\psi_{+,\mbf k'},
\enal
\eneq
where $3=2s+1$ and $S=0,1,2$ are the total spins of the Cooper pairs. Each component $\mca M_{SM_S}$ is a $3\times3$ irreducible representation of the $SU(2)$ group, which characterizes the total spin $S$ and $z$-component $M_S$ of a Cooper pair
\beeq
\psi_{+,\mbf k}^\dag\mca M_{SM_S}[\g(\psi_{-,-\mbf k}^\dag)^T].
\eneq
The normalization condition of the matrices is defined as
\beeq
\Tr\lf(\mca M_{SM_S}\mca M_{S'M_{S'}'}^\dag\ri) = 3\d_{SS'}\d_{M_SM_{S'}'}.
\eneq

The spatial potential may be treated in similar fashion \cite{Savary2017PRB}. Due to the full spherical symmetry, the orbital angular momentum $L$ and $z$-component $M_L$ are good quantum numbers for the spatial part. Therefore, the spatial potential $V(\mbf r-\mbf r')$ can be decomposed into a series of orbital modes with different angular momentums $L$'s. Consider the momentum space representation $V(\mbf k-\mbf k')$. The decomposition is performed by writing the potential as a series of orbital modes, which are represented by the spherical harmonics $Y_{LM_L}(\hat{\mbf k})$'s
\beeq
\beal
V(\mbf k-\mbf k')
= 4\pi
&
\sum_{LM_L}V_L(k,k') \\
&~\times
\lf[k^LY_{LM_L}(\hat{\mbf k})\ri]
\lf[k'^LY_{LM_L}^*(\hat{\mbf k}')\ri]
.
\enal
\eneq
The coefficients $V_L(k,k')$'s, which can be regarded as the coupling strengths, are isotropic functions which only depend on the magnitudes of momenta $k$ and $k'$
\beeq
\beal
V_L(k,k')
&= \fr{1}{4\pi(2L+1)k^Lk'^L}
\\
&\quad~\times
\sum_{M_L}\int_{\O\O'} V(\mbf k-\mbf k')Y_{LM_L}^*(\hat{\mbf k})Y_{LM_L}(\hat{\mbf k}').
\enal
\eneq
We assume that the coupling strengths $V_L$'s are constants, which amounts to keeping the leading (most relevant) terms in a Taylor expansion of a short range interaction. 

Thus far, we have decomposed the spin and spatial sectors in isolation. However, given the strong spin-orbit coupling in the noninteracting Hamiltonian, the good quantum numbers for the pairing channels are $L$, $S$, total angular momentum $J=L+S$, and $z$-component of total angular momentum $M_J$. Therefore, the interacting action Eq.~(\ref{EqIntAc}) is better expressed as a summation over different pairing channels labelled by the good quantum numbers of the full problem, i.e.
\beeq
S_\mrm{int}
= -\fr{1}{6\mca V}\sum_{SLJ}\int_\tau\sum_{\mbf k\mbf k'}V_L
(\vec\Pi_{J,\mbf k}^{SL})^\dag\cdot\vec\Pi_{J,\mbf k'}^{SL}
.
\eneq
Each operator
\beeq
(\vec\Pi_{J,\mbf k}^{SL})^\dag
=
\psi_{+,\mbf k}^\dag k^L\vec{\mca N}_J^{SL}[\g(\psi_{-,-\mbf k}^\dag)^T]
\eneq
creates a zero momentum Cooper pair with total angular momentum $J$ and $z$-component $M_J$ \cite{Venderbos}. The irreducible representation of total angular momentum $\vec{\mca N}_{J}^{SL}$ is an addition of the spin and orbital modes
\beeq
\beal
\mca N_{JM_J}^{SL}
&= \sum_{M_LM_S}\lf\la \lf.LS;M_LM_S\ri|LS;JM_J\ri\ra
\\
&\quad\times
\lf[\sqrt{4\pi}Y_{LM_L}(\hat{\mbf k})\ri]\mca M_{SM_S},
\enal
\eneq
where the summation with Clebsch-Gordan coefficients $\la LS;M_LM_S|LS;JM_J\ra$ obeys $M_L+M_S=M_J$. Notice that inversion symmetry is absent in our setup. Consequently, parity quantum number is not a good label for the pairing channels.

With the interactions decomposed into irreducible pairing channels of different angular momenta, the study of superconductivity in each channel $(L,S,J)$ may be performed by considering the action
\begin{align}
S^{(L,S,J)}
&= S_0+S_\mrm{int}^{(L,S,J)} \\
&\beal
= -\int_\tau
&
\Bigg[
\sum_{\mbf k}\Psi_{\mbf k}^\dag(\tau)\mca G_0^{-1}(\mbf k,\tau)\Psi_{\mbf k}(\tau) \\
&
+\fr{1}{6\mca V}\sum_{\mbf k\mbf k'}V_L
(\vec\Pi_{J,\mbf k}^{SL})^\dag\cdot\vec\Pi_{J,\mbf k'}^{SL}
\Bigg]
.
\enal
\label{eq:intchanact}
\end{align}
Due to Fermi statistics, the even-$S$ spin states must have even-$L$, while the odd-$S$ spin states must have odd-$L$. Different choice of quantum numbers $S$, $L$, and $J$ can lead to different types of superconductivity.

We now discuss the range of the momentum space summation. In BCS theory, the attractive interaction $-V$ works within the narrow energy domain $|\ve_{\mbf k}-\mu|<\o_D$, where $\o_D\ll v$ is the Debye frequency. The momentum space summations $\sum_{\mbf k}^{0,\pm}$ for the three bands are therefore restricted, and are different from each other in general. This feature can be characterized by the density of states $\nu^{0,\pm}(\ve)$ near chemical potential $\mu$
\beeq
\fr{1}{\mca V}\lf.\sum_{\mbf k}\ri.^{0,\pm}=\int_{\mu-\o_D}^{\mu+\o_D}d\ve\nu^{0,\pm}(\ve).
\eneq
A remarkable thing happens when the chemical potential is near zero $\mu\apx0$. While the densities of states $\nu^\pm(\ve)$ for linearly dispersing bands both vanish, the flat band acquires an extremely large density of states (in fact, a delta function at zero energy). As elucidated in the following sections, this abnormally large density of states can lead to striking departures from the conventional BCS theory.

\subsection{Bogoliubov-de Gennes Quasiparticle Spectrum}

Before embarking on the coherent path integral studies of pairing states, we analyze the properties of BdG quasiparticles in the framework of second quantization. The analysis will also be useful in the following path integral calculations.

With the analysis of pairing interactions in Sec.~\ref{sec:intpairchan}, the second quantized BCS Hamiltonian in the pairing channel $(L,S,J)$ can be determined as \cite{AltlandSimons,Coleman}
\beeq
\beal
H^{(L,S,J)} -\mu N
&= \sum_{\mbf k}
\sum_{a=\pm}c_{a,a\mbf k}^\dag\lf[v(a\mbf k)\cdot\mbf S-\mu\ri]c_{a,a\mbf k}
\\
&\quad-
\fr{1}{6\mca V}\sum_{\mbf k\mbf k'}V_L
(\vec\Pi_{J,\mbf k}^{SL})^\dag\cdot\vec\Pi_{J,\mbf k'}^{SL}
,
\enal
\eneq
where $(\vec\Pi_{J,\mbf k}^{SL})^\dag$ is the creation operator of a zero momentum Cooper pair with quantum numbers $(L,S,J,M_J)$ \cite{Venderbos}
\beeq
(\vec\Pi_{J,\mbf k}^{SL})^\dag
=
c_{+,\mbf k}^\dag k^L\vec{\mca N}_J^{SL}[\g(c_{-,-\mbf k}^\dag)^T]
.
\eneq
As in Sec. \ref{sec:intpairchan}, the momentum space summation is performed within the narrow energy range $|\ve_{\mbf k}-\mu|<\o_D$.

Define the gap funtion $\vec\D$, also known as the order parameter, and its adjoint $\vec{\bar\D}$ as the ground state expectation values of Cooper pair operators
\begin{align}
\vec\D
&= \fr{V_L}{6\mca V}\sum_{\mbf k}
\lf\la\O_s\ri|
\vec\Pi_{J,\mbf k}^{SL}
\lf|\O_s\ri\ra,\\
\vec{\bar\D}
&= \fr{V_L}{6\mca V}\sum_{\mbf k}
\lf\la\O_s\ri|
(\vec\Pi_{J,\mbf k}^{SL})^\dag
\lf|\O_s\ri\ra,
\end{align}
where $\ket{\O_s}$ is the BCS ground state in this channel. The quartic interaction term in the Hamiltonian can be replaced by the gap functions $\vec\D$ and $\vec{\bar\D}$
\begin{align}
H^{(L,S,J)}-\mu N
&= \sum_{\mbf k}\Bigg\{
\sum_{a=\pm}
c_{a,a\mbf k}^\dag\lf[v(a\mbf k)\cdot\mbf S-\mu\ri]c_{a,a\mbf k}
\nonumber\\
&\quad\quad\quad-
\vec{\bar\D}\cdot\vec\Pi_{J,\mbf k}^{SL}
-\vec\D\cdot(\vec\Pi_{J,\mbf k}^{SL})^\dag
\Bigg\}
\nonumber\\
&\quad+
\fr{6\mca V}{V_L}|\vec\D|^2
.
\end{align}
With similar treatment of the $(c_{-,-\mbf k})$-part in Sec.~\ref{SecAc}, the Hamiltonian can be expressed in the form of Bogoliubov-de Gennes (BdG) Hamiltonian
\beeq
H^{(L,S,J)}-\mu N = \sum_{\mbf k}\Psi_{\mbf k}^\dag\mca H_\mrm{BdG}^{(L,S,J)}(\mbf k)\Psi_{\mbf k}
-3\sum_{\mbf k}\mu+\fr{6\mca V}{V_L}|\vec\D|^2.
\eneq
The vector $\Psi_{\mbf k}$ is defined as the Nambu spinor
\beeq
\Psi_{\mbf k} = \lf(\bear{c}c_{+,\mbf k}\\\g(c_{-,-\mbf k}^\dag)^T\enar\ri),
\eneq
and $\mca H_\mrm{BdG}^{(L,S,J)}(\mbf k)$ is the matrix representation of BdG Hamiltonian
\beeq
\label{eq:bdgham}
\mca H_\mrm{BdG}^{(L,S,J)}(\mbf k)
= \lf(\bear{cc}
\mca H_0(\mbf k) & -\vec\D\cdot k^L\vec{\mca N}_J^{SL} \\
-\vec{\bar\D}\cdot\lf(k^L\vec{\mca N}_J^{SL}\ri)^\dag & -\mca H_0(\mbf k)
\enar\ri)
\eneq
in this channel. Notice that the diagnoal elements of $\mca H_\mrm{BdG}^{(L,S,J)}(\mbf k)$ are noninteracting Hamiltonians $\pm\mca H_0(\mbf k)$
\beeq
\label{eq:nonintham}
\mca H_0(\mbf k)
= v\mbf k\cdot\mbf S-\mu
=
\lf(\bear{ccc}
vk_3-\mu & \fr{1}{\sqrt2}vk_- & 0 \\
\fr{1}{\sqrt2}vk_+ & -\mu & \fr{1}{\sqrt2}vk_- \\
0 & \fr{1}{\sqrt2}vk_+ & -vk_3-\mu
\enar\ri)
\eneq
about the two band crossing points $\pm P$, where the momentum variables $k_\pm = k_x\pm ik_y$ are defined. The eigenvalues of $\mca H_0(\mbf k)$ are given by
\begin{align}
\xi_{\mbf k}^0 &= \ve_{\mbf k}^0-\mu = -\mu,\\
\xi_{\mbf k}^\pm &= \ve_{\mbf k}^\pm-\mu = \pm vk-\mu.
\end{align}

By diagonalizing the BdG Hamiltonian Eq.~(\ref{eq:bdgham}), the energies $\l_{\mbf k}^a$'s of BdG quasiparticles can be obtained. These BdG quasiparticle energies indicate various properties of the pairing state, including the energy gap and the pairing between noninteracting states about the two band crossing points $\pm P$. They also play important roles in the derivation of gap equation, which is the main topic of Sec.~\ref{sec:acttogapeq}.

\subsection{From Action to Gap Equation}
\label{sec:acttogapeq}

\subsubsection{Partition Function}

In order to probe the superconductivity in each irreducible pairing channel $(L,S,J)$, we study the quantum partition function of the system, which appears as a coherent path integral \cite{AltlandSimons,Coleman}
\beeq
Z^{(L,S,J)} = \int\prod_{a=\pm}\mca{D}\lf(\bar\psi_a,\psi_a\ri) e^{-S^{(L,S,J)}[\bar\psi,\psi]}
\eneq
over the fields $\psi_\pm$. The action $S^{(L,S,J)}$ is determined by the previous analysis of pairing interactions in Sec.~\ref{sec:intpairchan}, and is given by Eq.~(\ref{eq:intchanact}).

The quartic interaction in the action $S^{(L,S,J)}$ prohibits exact calculations of the path integral. To decouple the quartic interaction term, the Hubbard-Stratonovich transformation \cite{AltlandSimons,Coleman} is applied to the path integral. At mean field level, the transformation is performed by introducing the path integral of an auxiliary $M_J$-component bosonic field $\vec\D(\tau)$
\beeq
\beal
Z^{(L,S,J)}
&= \int\mca{D}\lf(\vec{\bar\D},\vec\D\ri)\int\prod_{a=\pm}\mca{D}\lf(\bar\psi_a,\psi_a\ri) \\
&\quad~
\exp\int_\tau\Bigg\{
-\fr{6\mca V}{V_L}|\vec\D|^2
-\sum_{\mbf k}
\Big[
-\Psi_{\mbf k}\mca G_0^{-1}(\mbf k)\Psi_{\mbf k} \\
&\quad\quad\quad\quad\quad
-\vec\D\cdot(\vec\Pi_{J,\mbf k}^{SL})^\dag
-\vec{\bar\D}\cdot\vec\Pi_{J,\mbf k}^{SL}
\Big]
\Bigg\},
\enal
\eneq
where the quartic interaction is replaced. The field $\vec\D$ is identified with the gap function of the superconducting state. Define the interacting inverse Gor'kov Green's function
\beeq
\mca G^{-1}(\mbf k,\tau) =
\lf(\bear{cc}
-\p_\tau-v\mbf k\cdot\mbf S+\mu & \vec\D\cdot k^L\vec{\mca N}_J^{SL} \\
\vec{\bar\D}\cdot(k^L\vec{\mca N}_J^{SL})^\dag & -\p_\tau+v\mbf k\cdot\mbf S-\mu
\enar\ri)
,
\eneq
which is related to the BdG Hamiltonian Eq.~(\ref{eq:bdgham}) by
\beeq
\mca G^{-1}(\mbf k,\tau) = -\p_\tau-\mca H_\mrm{BdG}^{(L,S,J)}(\mbf k).
\eneq
The partition function can be written as a path integral of the gap function $\vec\D$ and the Nambu spinor $\Psi_{\mbf k}$
\begin{align}
Z^{(L,S,J)}
&= \int\mca{D}\lf(\vec{\bar\D},\vec\D\ri)\int\mca{D}\lf(\Psi^\dag,\Psi\ri)
\nonumber\\
&\quad~
\exp\int_\tau\lf[-\fr{6\mca V}{V_L}|\vec\D|^2-\sum_{\mbf k}\lf(-\Psi_{\mbf k}^\dag\mca{G}^{-1}(\mbf k)\Psi_{\mbf k}\ri)\ri]
.
\end{align}
Integrating out the Nambu spinor field $\Psi_{\mbf k}$, the only remaining field in the path integral is the gap function $\vec\D$
\beeq
\label{eq:pathintgapfun}
Z^{(L,S,J)}
= \int\mca{D}\lf(\vec{\bar\D},\vec\D\ri)e^{-\b F^{(L,S,J)}[\vec{\bar\D},\vec\D]}.
\eneq
The free energy
\beeq
\label{eq:freeenergy}
F^{(L,S,J)}[\vec{\bar\D},\vec\D] = \fr{6\mca V}{\b V_L}\int_\tau|\vec\D|^2-\fr{1}{\b}\ln\det\mca{G}^{-1}
\eneq
includes a determinant of the inverse Gor'kov Green's function as the product of Nambu spinor path integral.

It is worth mentioning that although the quartic interaction in the  original action $S^{(L,S,J)}$ is decoupled, the path integral Eq.~(\ref{eq:pathintgapfun}) is still not exactly solvable. 
However, it serves as a useful starting point for investigations of superconductivity. For example, the expansion of free energy $F^{(L,S,J)}[\vec{\bar\D},\vec\D]$ Eq.~(\ref{eq:freeenergy}) with respect to the gap function $\vec\D$ near the critical temperature $T_c$ provides an approximate description of the phase transition between superconducting and normal states \cite{AltlandSimons}. One can also derive the mean field gap function $\vec\D$ and critical temperature $T_c$ from the gap equation, which describes the static solution of the path integral Eq.~(\ref{eq:pathintgapfun}). This analysis is presented in the following subsection.

\subsubsection{Gap Equation}

At mean field level, the gap function can be obtained from a self consistent gap equation, which in turn can be determined from the mean field free energy $F^{(L,S,J)}[\vec{\bar\D},\vec\D]$. Assume that the gap function is temporally static $\vec\D(\tau) = \vec\D$. The mean field free energy $F^{(L,S,J)}[\vec{\bar\D},\vec\D]$ takes the form
\beeq
F^{(L,S,J)}[\vec{\bar\D},\vec\D] = \fr{6\mca V}{V_L}|\vec\D|^2-\fr{1}{\b}\ln\det\mca{G}^{-1}.
\eneq
We apply the Fourier transform in temporal direction
\begin{gather}
\Psi_{\mbf k}(\tau) = \fr{1}{\sqrt\b}\sum_n\Psi_{\mbf kn}e^{-i\o_n\tau},
\\
\Psi_{\mbf kn} = \lf(\bear{c}\psi_{+,\mbf kn}\\\g(\psi_{-,-\mbf k-n}^\dag)^T\enar\ri)
,\\
\mca G^{-1}(\mbf k,\o_n) = i\o_n-\mca H_\mrm{BdG}^{(L,S,J)}(\mbf k),
\label{eq:relainvggfbdgham}
\end{gather}
where the Matsubara frequencies are $\o_n=(2n+1)\pi/\b$ with $n\in\mbb Z$ since the system is fermionic. This yields
\beeq
\label{eq:freeenergymf}
F^{(L,S,J)}[\vec{\bar\D},\vec\D] = \fr{6\mca V}{V_L}|\vec\D|^2-\fr{1}{\b}\sum_{\mbf kn}\ln\det\mca{G}^{-1}(\mbf k,n).
\eneq
The determinant of inverse Gor'kov Green's function can be evaluated by diagonalization. Due to the relation Eq.~(\ref{eq:relainvggfbdgham}) between inverse Gor'kov Green's function and BdG Hamiltonian, the result is determined by the energies $\l_{\mbf k}^a$'s of BdG quasiparticles
\beeq
\det\mca G^{-1}(\mbf k,n) = \prod_a\lf(i\o_n-\l_{\mbf k}^a\ri).
\eneq
The free energy Eq.~(\ref{eq:freeenergymf}) becomes a summation over BdG quasiparticle modes
\beeq
F^{(L,S,J)}[\vec{\bar\D},\vec\D] = \fr{6\mca V}{V_L}|\vec\D|^2-\fr{1}{\b}\sum_a\lf.\sum_{\mbf k}\ri.^a\sum_n\ln\lf(i\o_n-\l_{\mbf k}^a\ri).
\eneq

The gap equation can be determined by minimizing the free energy $F^{(L,S,J)}[\vec{\bar\D},\vec\D]$ with respect to $\vec\D$
\beeq
\fr{6\bar\D_{M_J}}{V_L}
= -\fr{1}{\b\mca V}\sum_a\lf.\sum_{\mbf k}\ri.^a\sum_n
\fr{1}{i\o_n-\l_{\mbf k}^a}\pd{\l_{\mbf k}^a}{\D_{M_J}}.
\eneq
We carry out the Matsubara frequency summation
\beeq
\label{EqMFS}
\fr{1}{\b}\sum_nh(i\o_n) = \oint\fr{dz}{2\pi i}f(z)h(z),
\eneq
where the Fermi function
\beeq
\label{eq:fermifunction}
f(z) = \fr{1}{e^{\b z}+1}
\eneq
provides simple poles $z_n=i\o_n$ with residues $-1/\b$. The integral contour surrounds the poles $z_n$'s on the imaginary axis in a clockwise way, and can be deformed into counterclockwise contour around the poles of $h(z)$. After the Matsubara frequency summation, the gap equation becomes
\beeq
\fr{6\bar\D_{M_J}}{V_L}
= -\sum_a\fr{1}{\mca V}\lf.\sum_{\mbf k}\ri.^a
f(\l_{\mbf k}^a)\pd{\l_{\mbf k}^a}{\D_{M_J}}.
\label{eq:gapeqf}
\eneq
Notice that the exact form of gap equation in each irreducible pairing channel depends explicitly on the BdG quasiparticle energies.

\subsubsection{Critical Temperature}

For a system which exhibits superconductivity in low temperature regime, there is a critical temperature $T_c$ which separates superconducting and normal states. When the temperature $T$ increases from $T=0$, the nonzero superconducting gap function $\vec\D$ usually decreases, and ultimately vanishes at the critical temperature $T=T_c$. Above the critical temperature, the gap function is always zero, implying that only the normal state exists in the high temperature regime.

The gap equation Eq.~(\ref{eq:gapeqf}) provides a way to calculate the critical temperature $T_c$ \cite{AltlandSimons,Coleman}. Normally, the right-hand side of the equation carries a factor $\bar\D_{M_J}$, which eliminates the same factor on the left-hand side. This exclude the trivial solution $\vec\D=0$ from the equation, and only the nontrivial solutions remain. By taking $\vec\D=0$ in the remaining equation, the critical temperature $T=T_c$ at which the nontrivial solution, hence the superconductivity, vanishes can be solved.

This concludes the general setup of the basic BCS problem in the three band crossing setting.

\section{Superconductivity of spin one fermions}
\label{sec:scspinonefermion}

In this section, we analyze the superconducting states in different irreducible pairing channels. For simplicity, we study only the single-component channels with $J=0$, including the $s$-wave singlet channel $(L,S,J)=(0,0,0)$, the $p$-wave triplet channel $(L,S,J)=(1,1,0)$, and the $d$-wave quintet channel $(L,S,J) = (2,2,0)$. The zero total angular momentum $J=0$ implies that the vector of irreducible representation for pairing has only one component $\mca N_{00}^{LL}$. Correspondingly, the gap function is a scalar $\D$.

For each $J=0$ pairing channel, one can verify that the off-diagonal pairing representation $k^L\mca N_{00}^{LL}$ in BdG Hamiltonian Eq.~(\ref{eq:bdgham}) is Hermitian. This implies that its eigenvalues $\z_{\mbf k}$'s are real. In addition, the pairing representation $k^L\mca N_{00}^{LL}$ commutes with the diagonal noninteracting Hamiltonian $\mca H_0(\mbf k)$ Eq.~(\ref{eq:nonintham}). Therefore, each eigenvalue $\z_{\mbf k}^a$ corresponds to an eigenstate $\ket{u_{\mbf k}^a}$ of $\mca H_0(\mbf k)$. The two features provide a significant simplification of BdG Hamiltoinan Eq.~(\ref{eq:bdgham}). By choosing the eigenbasis $\{\ket{u_{\mbf k}^0},\ket{u_{\mbf k}^-},\ket{u_{\mbf k}^+}\}$ of $\mca H_0(\mbf k)$ in both $(c_{+,\mbf k})$- and $[\g(c_{-,-\mbf k}^\dag)^T]$-sections, the BdG Hamiltonian becomes a composition of diagonal blocks
\beeq
\beal
&\mca  H_\mrm{BdG}^{(L,L,0)}(\mbf k) \\
&=
\lf(\bear{cccccc}
\xi_{\mbf k}^0&0&0&-\D\z_{\mbf k}^0&0&0\\
0&\xi_{\mbf k}^-&0&0&-\D\z_{\mbf k}^-&0\\
0&0&\xi_{\mbf k}^+&0&0&-\D\z_{\mbf k}^+\\
-\bar\D\z_{\mbf k}^0&0&0&-\xi_{\mbf k}^0&0&0\\
0&-\bar\D\z_{\mbf k}^-&0&0&-\xi_{\mbf k}^-&0\\
0&0&-\bar\D\z_{\mbf k}^+&0&0&-\xi_{\mbf k}^+
\enar\ri).
\enal
\eneq
A reordering of basis states further transforms the matrix into a block-diagonal form
\beeq
\label{eq:bdgblkdiag}
\mca H_\mrm{BdG}^{(L,L,0)}(\mbf k)
= \bigoplus_{a=0,\pm}\mca H_\mrm{BdG}^{(L,L,0),a}(\mbf k).
\eneq
Each block
\beeq
\label{eq:bdghamnoninteigmode}
\mca H_\mrm{BdG}^{(L,L,0),a}(\mbf k) =
\lf(\bear{cc}
\xi_{\mbf k}^a & -\D\z_{\mbf k}^a\\
-\bar\D\z_{\mbf k}^a&-\xi_{\mbf k}^a
\enar\ri)
\eneq
is the BdG Hamiltonian in the $\ket{u_{\mbf k}^a}$-section, which describes the BCS physics in single band theory \cite{AltlandSimons,Coleman}. The BdG Hamiltonian is now decomposed into independent sections, each of which results from a single type of noninteracting eigenstate $\ket{u_{\mbf k}^a}$. The feasibility of such decomposition indicates that Cooper pairing only happens between the bands about $P$ and $-P$ with the same band index. By diagonalizing the BdG Hamiltonian $\mca H_\mrm{BdG}^{(L,L,0),a}(\mbf k)$ in each $\ket{u_{\mbf k}^a}$-section, the BdG quasiparticle energies are evaluated as
\beeq
\label{eq:bdgqpe}
\l_{\mbf k}^{a,\pm} = \pm\sqrt{|\D|^2(\z_{\mbf k}^a)^2+\lf(\xi_{\mbf k}^a\ri)^2},\quad
a=0,\pm.
\eneq
Different irreducible pairing channels provide different sets of eigenvalues $\z_{\mbf k}^a$'s, which leads to different energy spectrums of BdG quasiparticles.

With the BdG quasiparticle energies, the gap equation Eq.~(\ref{eq:gapeqf}) becomes
\beeq
\fr{6\bar\D}{V_L}
= -\sum_{a=0,\pm}\fr{\bar\D}{\mca V}\lf.\sum_{\mbf k}\ri.^a\sum_{b=\pm}
\fr{b(\z_{\mbf k}^a)^2 f(b\sqrt{|\D|^2[\z_{\mbf k}^a]^2+[\xi_{\mbf k}^a]^2})}{2\sqrt{|\D|^2(\z_{\mbf k}^a)^2+(\xi_{\mbf k}^a)^2}}
.
\eneq
The equation supports a trivial solution $\D=0$ which corresponds to the normal state. Our interest lies in the nontrivial solutions $\D\neq0$. Since the Fermi function $f(z)$ Eq.~(\ref{eq:fermifunction}) satisfies
\beeq
f(z)-f(-z) = -\tanh\fr{\b}{2}z,
\eneq
the gap equation is equivalent to a summation over hyperbolic tangent functions in different modes
\beeq
\label{eq:gapeq}
\fr{6}{V_L}
= \sum_{a=0,\pm}\fr{1}{\mca V}\lf.\sum_{\mbf k}\ri.^a
\fr{(\z_{\mbf k}^a)^2\tanh\fr{\b}{2}\sqrt{|\D|^2(\z_{\mbf k}^a)^2+(\xi_{\mbf k}^a)^2}}{2\sqrt{|\D|^2(\z_{\mbf k}^a)^2+(\xi_{\mbf k}^a)^2}}
.
\eneq
The calculation of gap equation is significantly simplified from the original manipulation of $6\times6$ matrices. Once the eigenvalues $\z_{\mbf k}^a$'s of pairing irreducible representation $\D k^L\mca N_{00}^{LL}$ are obtained, the gap equation in the pairing channel can be determined directly.

\subsection{$s$-wave Pairing}
\label{sec:swp}

As a starting point, we investigate the superconductivity in $s$-wave singlet pairing channel $(L,S,J)=(0,0,0)$. Since the orbital and spin parts are both trivial, the irreducible representation $\mca N_{00}^{00}$ in the Cooper pair is an identity matrix
\beeq
\mca N_{00}^{00} = 1.
\eneq
The only eigenvalue of pairing representation $\mca N_{00}^{00}$ is
\beeq
\z_{\mbf k} = 1.
\eneq

\subsubsection{Bogoliubov-de Gennes Quasiparticles}

\begin{figure}[t]
\includegraphics[scale=1]{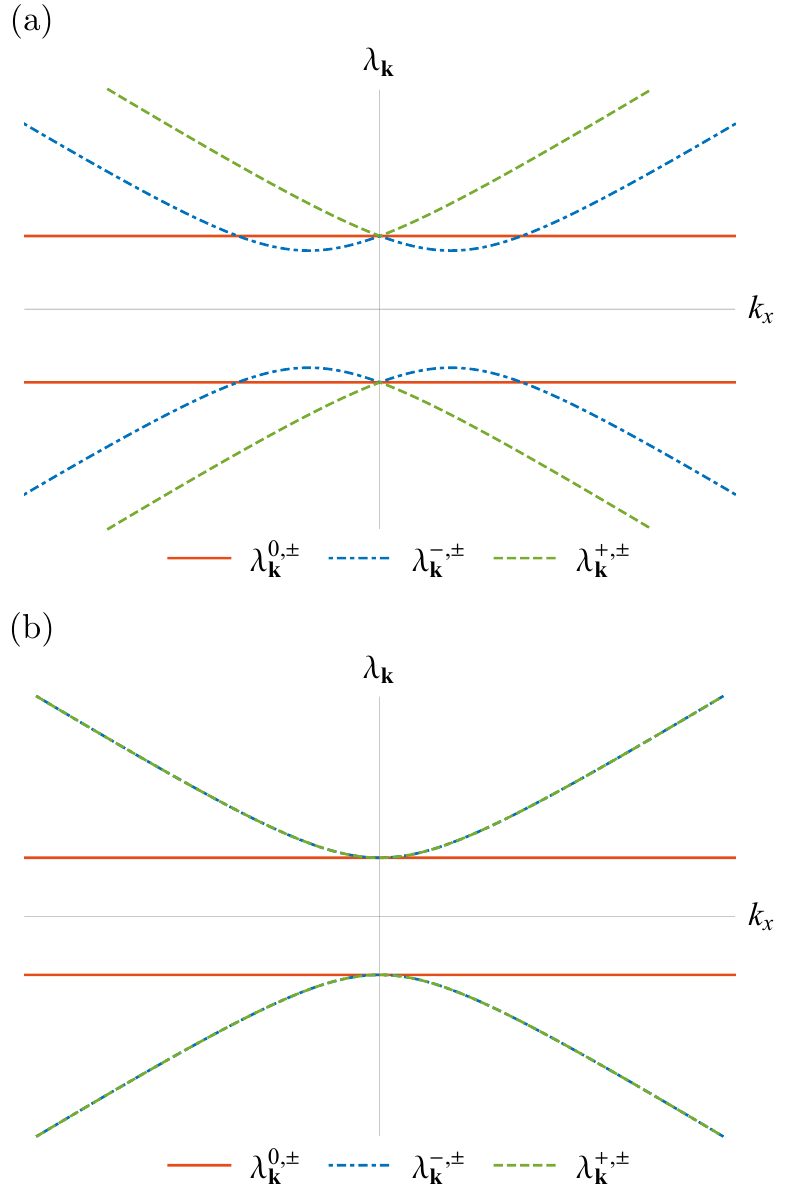}
\caption{\label{fig:swpbdg} Schematic illustrations of the BdG quasiparticle energy spectrum for (a) $\o_D>\mu>0$ and (b) $\mu=0$ in the $s$-wave singlet $(0,0,0)$ pairing channel. When $\mu$ goes across zero, the bands of $\l_{\mbf k}^-$ and $\l_{\mbf k}^+$ coincide and then switch with each other. The spectrum is fully gapped in either case, with energy gap $2|\D|$.}
\end{figure}

The energies of BdG quasiparticles are read off from the general form Eq.~(\ref{eq:bdgqpe})
\beeq
\l_{\mbf k}^{a,\pm} = \pm\sqrt{|\D|^2+\lf(\xi_{\mbf k}^a\ri)^2},\quad
a=0,\pm.
\eneq
An observation of energy spectrum Fig.~\ref{fig:swpbdg} indicates that the system is fully gapped. The minimal gap between positive and negative BdG quasiparticle bands is determined by the gap function $2|\D|$. Notice that the pairing between flat bands about the band crossing points $\pm P$ brings about the two BdG quasiparticle bands with energies $\l_{\mbf k}^{0,\pm}$. As indicated by the following calculation of gap equation, the existence of such flat band pairing can lead to novel superconductivity with enhanced gap function $\D$ and higher critical temperature $T_c$.

\subsubsection{Gap Function and Critical Temperature}

The gap equation Eq.~(\ref{eq:gapeq}) for $s$-wave singlet pairing channel takes the form
\beeq
\label{eq:gapeqswpini}
\fr{6}{V_0}
= \sum_{a=0,\pm}\fr{1}{\mca V}\lf.\sum_{\mbf k}\ri.^a
\fr{\tanh\fr{\b}{2}\sqrt{|\D|^2+(\xi_{\mbf k}^a)^2}}{2\sqrt{|\D|^2+(\xi_{\mbf k}^a)^2}}
.
\eneq
Each term in the sum results from the coupling between a single type of noninteracting band about the band crossing points $\pm P$.

We are particularly interested in the small chemical potential regime $|\mu|<\o_D$, where the Fermi level is close to the band crossing points $\pm P$.  Since the densities of states $\nu^\pm(\ve)$ of linearly dispersing bands are infinitesimal near $\ve=0$, the contribution from these two modes are negligible. In contrast to the linearly dispersing bands, the whole flat band is covered by the momentum space summation $\sum_{\mbf k}^0$, which enhances pairing. The dominating contribution from flat band pairing suggests a simplified form of gap equation
\beeq
\label{eq:gapeqswpmomspasum}
\fr{6}{V_0}
= \fr{1}{\mca V}\lf.\sum_{\mbf k}\ri.^0\fr{\tanh\fr{\b}{2}\sqrt{|\D|^2+\mu^2}}{2\sqrt{|\D|^2+\mu^2}}.
\eneq
Assume that the flat band exists within a spherical momentum space domain about each band crossing point. For the convenience of calculations, we assume a momentum cutoff $k=\L$ for the spherical domain, and rewrite the momentum space summation as a continous momentum integral
\beeq
\label{eq:momspasumtoint}
\fr{1}{\mca V}\lf.\sum_{\mbf k}\ri.^0 = \fr{1}{2\pi^2}\int_0^\L dkk^2.
\eneq
The integral version of gap equation reads
\beeq
\label{eq:gapeqswp}
\fr{6}{V_0}
= \til{\mca V}_\mrm{FB}\fr{\tanh\fr{\b}{2}\sqrt{|\D|^2+\mu^2}}{2\sqrt{|\D|^2+\mu^2}},
\eneq
where the volume of flat band domain $\til{\mca V}_\mrm{FB}=\L^3/6\pi^2$ characterizes the density of states $\nu^0(\ve)=\til{\mca V}_\mrm{FB}\d(\ve)$. The behavior of gap function can be probed from the examination of this equation.

\begin{figure}[b]
\includegraphics[scale=1]{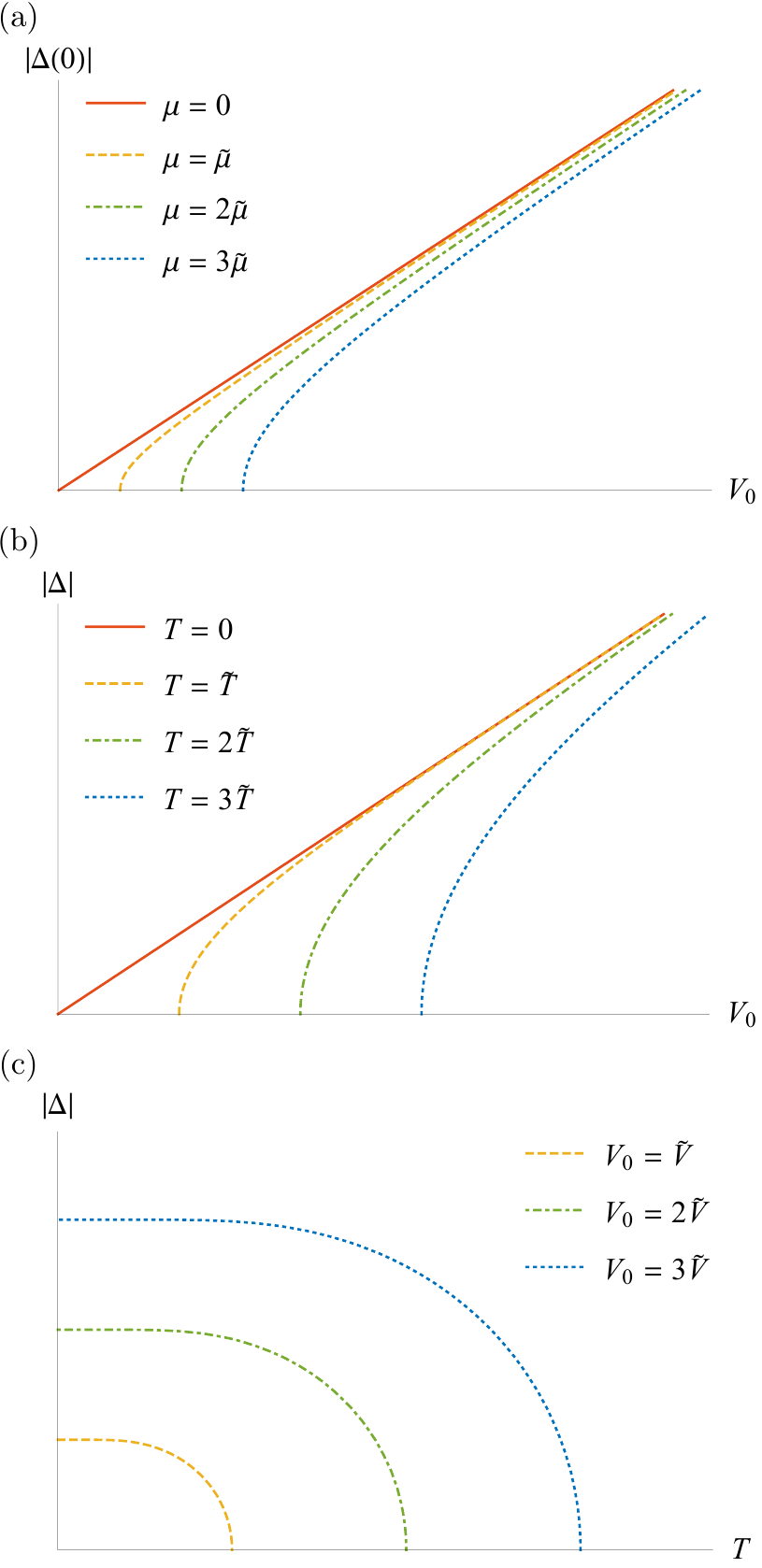}
\caption{\label{fig:swpgf} Numerically obtained gap function in the $s$-wave singlet pairing channel.  $\til T$, $\til V$, and $\til \mu$ are positive constants. (a) At $T=0$, gap function $\D(0)$ is proportional to coupling strength when $\mu=0$, and is suppressed by finite chemical potentials $\mu$'s. (b) When $\mu=0$, the gap function $\D$ is proportional to coupling strength $V_0$ at $T=0$, and is suppressed when $T>0$. (c) A critical temperature $T=T_c$ marks the full elimination of gap function $\D$.}
\end{figure}

In the zero temperature limit $\b\rar\infty$, the hyperbolic tangent function goes to $1$. The gap equation reduces to
\beeq
\fr{6}{V_0} = \til{\mca V}_\mrm{FB}\fr{1}{2\sqrt{|\D(0)|^2+\mu^2}},
\eneq
where $\D(0)$ denotes the zero temperature gap function. The solution of this equation
\beeq
\label{eq:gapfunswp}
\lf|\D(0)\ri| = \sqrt{\lf(\fr{V_0\til{\mca V}_\mrm{FB}}{12}\ri)^2-\mu^2}
\eneq
describes the dependence of zero temperature gap function $\D(0)$ on the coupling strength $V_0$ and the chemical potential $\mu$. A remarkable feature shows up when the Fermi level lies exactly at the band crossing points $\mu=0$. The zero temperature gap function $\D(0)$ acquires a linear dependence on the coupling strength $V_0$ [Fig.~\ref{fig:swpgf}(a)]
\beeq
\label{eq:lingapfunswp}
\lf|\D(0)\ri| = \fr{1}{12}V_0\til{\mca V}_\mrm{FB},
\eneq
unlike the conventional BCS problem where pairing strength is exponentially small in coupling strength \cite{AltlandSimons,Coleman}. This linear scaling of gap function is a result of the extremely large density of states of the flat band $\nu^0(0)$, which enhances the growth of gap function with increasing coupling strength $V_0$ (similar to Ref.~\onlinecite{volovik, Uchoa2013PRL}). It is therefore suggested that three band crossings may support unusually strong superconductivity. Furthermore, doping {\it suppresses} pairing, since it moves the Fermi surface away from the energy with large density of states. (While the above equation suggests absence of superconductivity for infinitesimal coupling when $\mu \neq 0$, bear in mind that in deriving the above equation we have neglected the linearly dispersing bands. The equation above is thus only valid for $V_0\til{\mca V}_\mrm{FB}/12 > |\mu|$, when the interaction can access the flat band. When this condition is violated, we need to retain the linearly dispersing bands in the analysis. This will yield superconductivity with conventional BCS scaling, where the density of states goes as $\mu^2$.)

When the temperature is finite, thermal fluctuations can lead to the suppression of superconductivity. This thermal effect can also be observed from the gap equation Eq.~(\ref{eq:gapeqswp}). The hyperbolic tangent function decreases from its zero temperature value $1$ as the temperature increases. Correspondingly, the denominator becomes smaller in order to solve the gap equation. The observation indicates that the gap function decreases with increasing temperature [Fig.~\ref{fig:swpgf}(b) and Fig.~\ref{fig:swpgf}(c)], and ultimately vanishes at the critical temperature $T=T_c$. This amounts to the phase transition between superconducting and normal states.

The critical temperature $T=T_c$ is calculated by taking $|\D|\rightarrow0$ in the gap equation Eq.~(\ref{eq:gapeqswp})
\beeq
\fr{6}{V_0}=\til{\mca V}_\mrm{FB}\fr{\tanh\fr{\b_c}{2}|\mu|}{2|\mu|},
\eneq
which provides the solution
\beeq
\label{eq:critemswp}
T_c = \fr{|\mu|}{2}\lf[\tanh^{-1}\lf(\fr{12|\mu|}{V_0\til{\mca V}_\mrm{FB}}\ri)\ri]^{-1}.
\eneq
As $\mu=0$, the critical temperature scales linearly as the gap function does Eq.~(\ref{eq:lingapfunswp})
\beeq
\label{eq:lincritemswp}
T_c = \fr{1}{24}V_0\til{\mca V}_\mrm{FB},
\eneq
unlike the exponentially small scaling of normal BCS superconductors \cite{AltlandSimons,Coleman}. Therefore, superconductivity in three band crossings may exhibit unusually high critical temperatures. Notice that the ration between zero temperature gap function $|\D(0)|$ and critical temperature is
\beeq
\label{eq:ratioztgftc}
\fr{2|\D(0)|}{T_c} = 4,
\eneq
which is different from the value $2|\D(0)|/T_c\apx3.53$ in conventional BCS superconductivity \cite{AltlandSimons, Coleman}.


It is worth mentioning that the results obtained in this section assume a perfect flat band. The effect of perturbations that give the flat band nonzero band curvature is discussed in Sec.~\ref{sec:bandcurvature}. Our discussion has also assumed that $|\mu| < \omega_D$. When this condition is violated, the interaction cannot access the flat band, no matter how strong it may be. In this case the pairing is dominated by the linearly dispersing bands (which were neglected in the preceding analysis). Assume that the Fermi level crosses one of the linearly dispersing bands $a\mu>0$. The gap equation becomes
\beeq
\fr{6}{V_0}
= \fr{1}{\mca V}\lf.\sum_{\mbf k}\ri.^a
\fr{\tanh\fr{\b}{2}\sqrt{|\D|^2+(\xi_{\mbf k}^a)^2}}{2\sqrt{|\D|^2+(\xi_{\mbf k}^a)^2}}
.
\eneq
The potential superconductivity in this regime acquires the conventional behavior. In the weak coupling regime, the gap function $\D$ and the critical temperature $T_c$ both acquire the exponentially small scaling in conventional BCS theory \cite{AltlandSimons, Coleman}
\beeq
\label{eq:expscaswp}
\lf|\D(0)\ri|,T_c
\sim\o_D\exp\lf(-\fr{6}{V_0\nu^a(\mu)}\ri)
,
\eneq
with $\nu^a(\mu) \sim \mu^2$. Notice that the scalings in the regimes $|\mu|<\o_D$ and $|\mu|>\o_D$ are remarkably different. Due to this significant difference, discontinuities in the scalings occur when the Fermi level crosses the Debye frequency energy scale $|\mu|=\o_D$. Such discontinuities actually result from the hard cutoff in the applicable energy domain of attractive interaction $|\ve-\mu|<\o_D$. By relaxing the hard cutoff to a soft one, the discontinuities in scalings can be smeared.


\subsection{$p$-wave Pairing}

In addition to the $s$-wave singlet pairing channel, we also expand our investigations to the irreducible pairing channels with higher orbital angular momenta. One of the channels we study is the $p$-wave triplet pairing channel $(L,S,J)=(1,1,0)$. The vector $\vec{\mca M}_1$ for $3\times3$ irreducible representations of $S=1$ is
\begin{align}
\mca M_{11}
&= -\fr{\sqrt3}{2}S_+
= \sqrt{\fr{3}{2}}\lf(\bear{ccc}0&-1&0\\0&0&-1\\0&0&0\enar\ri),\\
\mca M_{10}
&= \sqrt{\fr{3}{2}}S_z
= \sqrt{\fr{3}{2}}\lf(\bear{ccc}1&0&0\\0&0&0\\0&0&-1\enar\ri),\\
\mca M_{1-1}
&= \fr{\sqrt3}{2}S_-
= \sqrt{\fr{3}{2}}\lf(\bear{ccc}0&0&0\\1&0&0\\0&1&0\enar\ri),
\end{align}
and the orbital spherical harmonics for $L=1$ are
\beeq
\sqrt{4\pi}Y_{1\pm 1}(\hat{\mbf k}) = \mp\sqrt{\fr{3}{2}}\hat k_\pm,\quad
\sqrt{4\pi}Y_{10}(\hat{\mbf k}) = \sqrt3\hat k_z.
\eneq
To obtain the irreducible representation $\mca N_{00}^{11}$ for total angular momentum of Cooper pair, the spin irreducible representations $\mca M_{1M_S}$'s and the orbital spherical harmonics $Y_{1M_L}(\hat{\mbf k})$'s are added
\begin{align}
\mca N_{00}^{11}
&= \sqrt{\fr{4\pi}{3}}
\Big[
Y_{11}(\hat{\mbf k})\mca M_{1-1}-Y_{10}(\hat{\mbf k})\mca M_{10}+Y_{1-1}(\hat{\mbf k})\mca M_{11}
\Big]
\\
&= 
\lf(\bear{ccc}
-\sqrt{\fr{3}{2}}\hat k_z&-\fr{\sqrt3}{2}\hat k_-&0\\
-\fr{\sqrt3}{2}\hat k_+ & 0 & -\fr{\sqrt3}{2}\hat k_- \\
0 & -\fr{\sqrt3}{2}\hat k_+ & \sqrt{\fr{3}{2}}\hat k_z
\enar\ri).
\end{align}
The pairing representation $k\mca N_{00}^{11}$ takes the $\mbf k\cdot\mbf S$ form as the noninteracting Hamiltonian $\mca H_0(\mbf k)$ Eq.~(\ref{eq:nonintham}) does. This implies that $k\mca N_{00}^{11}$ commutes with $\mca H^0(\mbf k)$. The eigenvalues of $k\mca N_{00}^{11}$ are
\beeq
\z_{\mbf k}^0 = 0,\quad
\z_{\mbf k}^\pm = \mp\sqrt{\fr{3}{2}}k
.
\eneq

\subsubsection{Bogoliubov-de Gennes Quasiparticles}

\begin{figure}[b]
\includegraphics[scale=1]{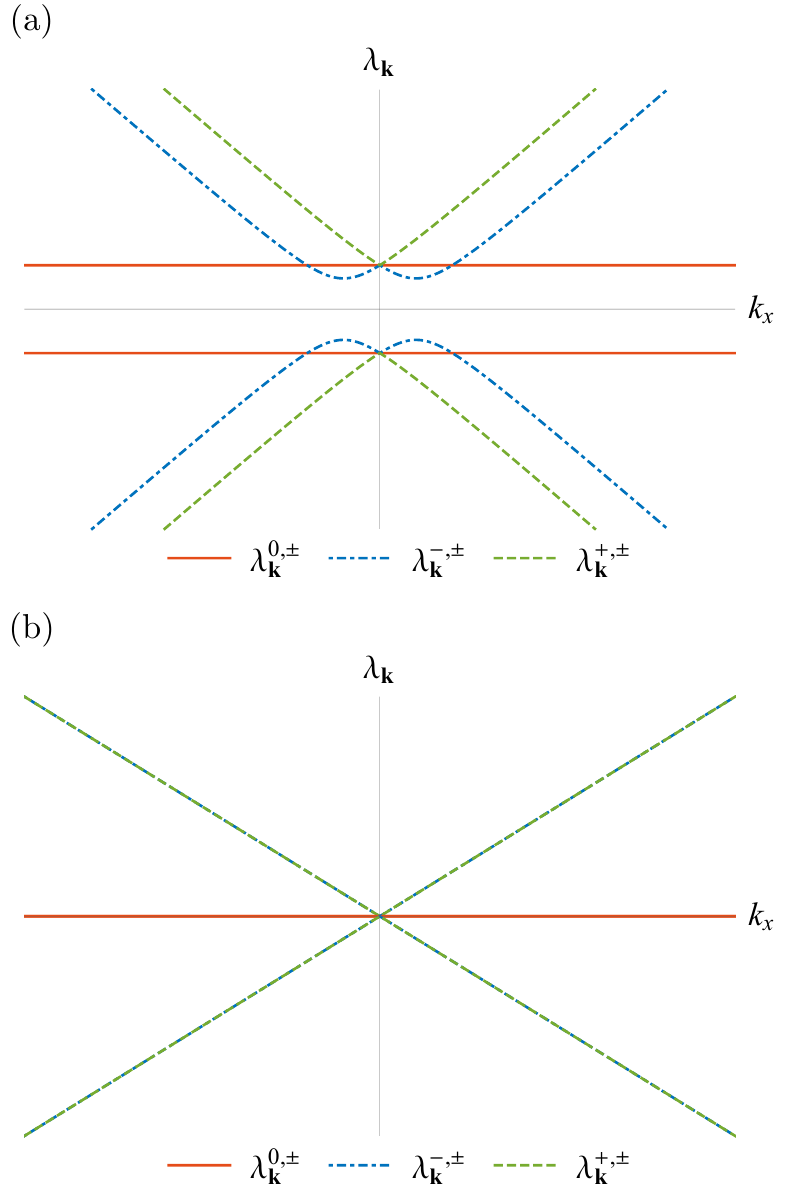}
\caption{\label{FigpWPBdG} Schematic illustration of the BdG quasiparticle energy spectrum for (a) $\o_D>\mu>0$ and (b) $\mu=0$ in the $p$-wave triplet $(1,1,0)$ pairing channel. The system is fully gapped when the Fermi level is away from band crossing points $\mu\neq0$. As $\mu$ goes across zero, the bands of $\l_{\mbf k}^-$ and $\l_{\mbf k}^+$ coincide and then switch with each other. Notice that the system becomes gapless when $\mu=0$. The absence of flat band pairing and the vanishing of the density of states of the linearly dispersing bands at $\mu=0$ implies the vanishing of the gap function $\D=0$ at $\mu=0$.}
\end{figure}

With the general form Eq.~(\ref{eq:bdgqpe}), the BdG quasiparticle energies in this channel are determined with $\z_{\mbf k}^a$'s
\begin{align}
\l_{\mbf k}^{0,\pm} &= \pm\lf|\xi_{\mbf k}^0\ri|,\\
\l_{\mbf k}^{a,\pm} &= \pm\sqrt{\fr{3}{2}k^2|\D|^2+\lf(\xi_{\mbf k}^a\ri)^2},\quad
a=\pm.
\end{align}
The energy spectrum is demonstrated in Fig.~\ref{FigpWPBdG}. Notice the absence of change in flat band energies, which indicates that the $p$-wave pairing, hence the superconductivity, does not affect or `benefit from' the flat band. The idleness of Cooper pairing between flat bands follows from the fact that flat band states are the zero modes of the operator $\mbf k\cdot\mbf S$, which describes both $p$-wave pairing representation $k\mca N_{00}^{11}$ and noninteracting Hamiltonian $\mca H_0(\mbf k)$. Without the enhancement of pairing from the extremely large density of states of flat band $\nu^0(0)$, the novel superconductivity with linear scaling in coupling strength is not supported in this channel.

\subsubsection{Gap Function and Critical Temperature}

The gap equation is determined from the general form for $J=0$ irreducible pairing channels Eq.~(\ref{eq:gapeq})
\beeq
\label{eq:gapeqpwp}
\fr{6}{V_1}
= \sum_{a=\pm}\fr{1}{\mca V}\lf.\sum_{\mbf k}\ri.^a\fr{3k^2\tanh\fr{\b}{2}\sqrt{3k^2|\D|^2/2+(\xi_{\mbf k}^a)^2}}{4\sqrt{3k^2|\D|^2/2+(\xi_{\mbf k}^a)^2}}
.
\eneq
Due to the lack of gap function $\D$ in the BdG quasiparticle energies $\l_{\mbf k}^{0,\pm}$, there is no flat band term in the gap equation. The absence of flat band contribution to gap equation justifies our argument that the flat band does not involve in the superconductivity in the $p$-wave triplet pairing channel. Since only the linearly dispersing bands contribute to the pairing state, the $p$-wave triplet pairing channel supports the conventional $p$-wave BCS superconductivity \cite{AltlandSimons, Coleman}. The gap function $\D$ and the critical temperature $T_c$ are characterized by exponentially small scaling, which are similar to Eq.~(\ref{eq:expscaswp}).

When the Fermi level is near the band crossing points $\mu\apx0$, the densities of states of both bands vanishes as $\nu^a(\mu) \sim \mu^2$. The critical temperature thus vanishes as $\ln T_c \sim -1/\mu^2$. Therefore, $p$-wave triplet pairing is strongly disfavored when the Fermi level is close to the band crossing points $\pm P$.

\subsection{$d$-wave Pairing}

The other higher angular momentum irreducible pairing channel we consider is the $d$-wave quintet pairing channel $(L,S,J)=(2,2,0)$. The vector $\vec{\mca M}_2$ of the $3\times3$ irreducible representations for $S=2$ is
\begin{align}
\mca M_{22}
&= \fr{\sqrt3}{2}S_+^2
= \sqrt3\lf(\bear{ccc}0&0&1\\0&0&0\\0&0&0\enar\ri),
\\
\mca M_{21}
&= -\fr{\sqrt3}{2}\lf\{S_+,S_z\ri\}
= \sqrt{\fr{3}{2}}\lf(\bear{ccc}0&-1&0\\0&0&1\\0&0&0\enar\ri),
\\
\mca M_{20}
&= \fr{1}{\sqrt2}\lf[3\lf(S^z\ri)^2-\mbf S^2\ri]
= \fr{1}{\sqrt2}\lf(\bear{ccc}1&0&0\\0&-2&0\\0&0&1\enar\ri),
\\
\mca M_{2-1}
&= \fr{\sqrt3}{2}\lf\{S_-,S_z\ri\}
= \sqrt{\fr{3}{2}}\lf(\bear{ccc}0&0&0\\1&0&0\\0&-1&0\enar\ri),
\\
\mca M_{2-2}
&= \fr{\sqrt3}{2}S_-^2
= \sqrt3\lf(\bear{ccc}0&0&0\\0&0&0\\1&0&0\enar\ri),
\end{align}
and the orbital spherical harmonics for $L=2$ are
\beeq
\bega
\sqrt{4\pi}Y_{2\pm 2}(\hat{\mbf k}) = \sqrt{\fr{15}{8}}\hat k_\pm^2,\quad
\sqrt{4\pi}Y_{2\pm 1}(\hat{\mbf k}) = \mp\sqrt{\fr{15}{2}}\hat k_\pm\hat k_z,
\\
\sqrt{4\pi}Y_{20}(\hat{\mbf k}) = \sqrt{\fr{5}{4}}\lf(3\hat k_z^2-1\ri).
\enga
\eneq
By adding the spin irreducible representations $\mca M_{2M_S}$'s and the orbital spherical harmonics $Y_{2M_L}(\hat{\mbf k})$'s, the irreducible representation $\mca N_{00}^{22}$ for total angular momentum of Cooper pair is derived
\begin{align}
\mca N_{00}^{22}
&= \sqrt{\fr{4\pi}{5}}
\Big[
Y_{22}(\hat{\mbf k})\mca M_{2-2}-Y_{21}(\hat{\mbf k})\mca M_{2-1}+Y_{20}(\hat{\mbf k})\mca M_{20}
\nonumber\\
&\quad\quad\quad~~
-Y_{2-1}(\hat{\mbf k})\mca M_{21}+Y_{2-2}(\hat{\mbf k})\mca M_{22}
\Big]
\\
&= 
\lf(\bear{ccc}
\fr{1}{2\sqrt2}(3\hat k_z^2-1) & \fr{3}{2}\hat k_-\hat k_z & \fr{3}{2\sqrt2}\hat k_-^2 \\
\fr{3}{2}\hat k_+\hat k_z & -\fr{1}{\sqrt2}(3\hat k_z^2-1) & -\fr{3}{2}\hat k_-\hat k_z \\
\fr{3}{2\sqrt2}\hat k_+^2 & -\fr{3}{2}\hat k_+\hat k_z & \fr{1}{2\sqrt2}(3\hat k_z^2-1)
\enar\ri).
\end{align}
It can be verified that the pairing representation $k^2\mca{N}_{00}^{22}$ is Hermitian and commutes with the noninteracting Hamiltonian $\mca H_0(\mbf k)$ Eq.~(\ref{eq:nonintham}). The eigenvalues of $k^2\mca{N}_{00}^{22}$ are
\beeq
\z_{\mbf k}^0 = -\sqrt2k^2,\quad
\z_{\mbf k}^\pm = \fr{k^2}{\sqrt2}.
\eneq

\subsubsection{Bogoliubov-de Gennes Quasiparticles}

\begin{figure}[b]
\includegraphics[scale=1]{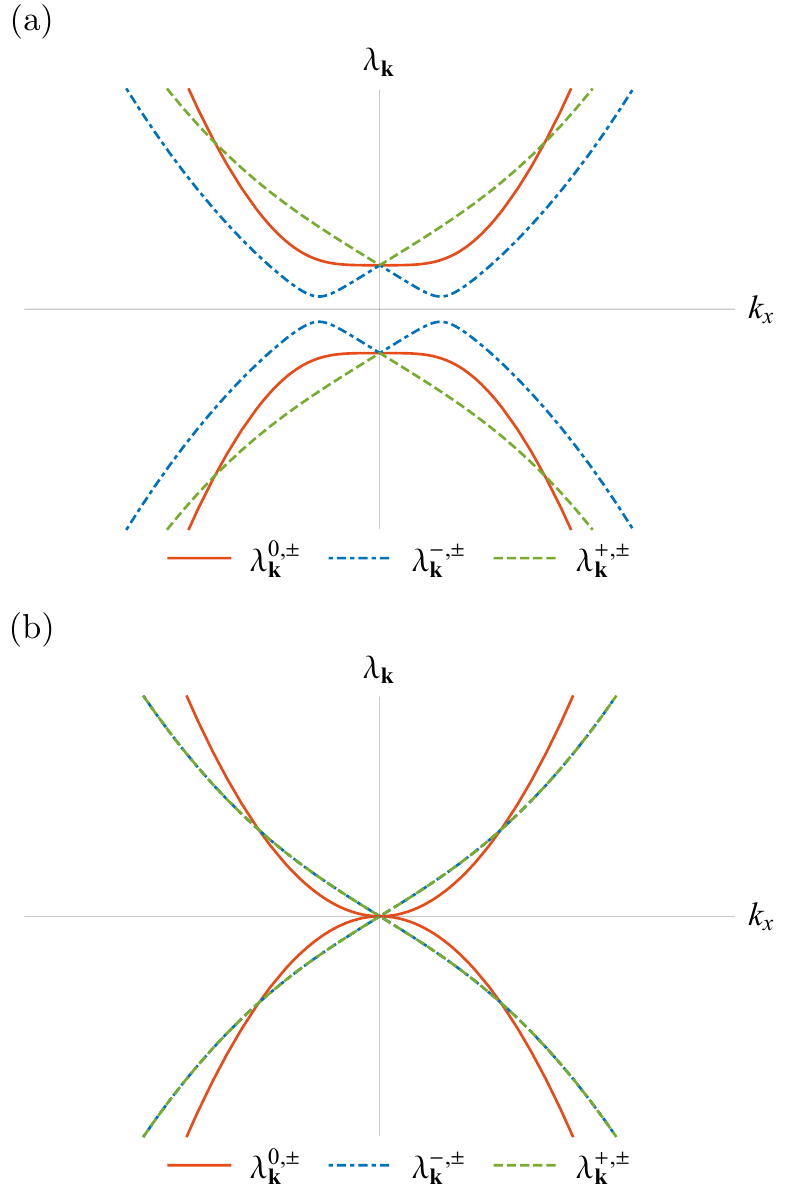}
\caption{\label{FigdWPBdG} Schematic illustration of the BdG quasiparticle spectrum for (a) $\o_D>\mu>0$ and (b) $\mu=0$ in the $d$-wave quintet $(2,2,0)$ pairing channel. The spectrum is fully gapped when $\mu\neq0$. As $\mu$ goes across zero, the bands of $\l_{\mbf k}^-$ and $\l_{\mbf k}^+$ coincide and then switch with each other. A nodal point which separates the positive and negative energy bands emerges at $\mbf k=0$ when $\mu=0$.}
\end{figure}

We obtain the BdG quasiparticle energies by inserting $\z_{\mbf k}^a$'s into the general form Eq.~(\ref{eq:bdgqpe})
\begin{align}
\l_{\mbf k}^{0,\pm} &= \pm\sqrt{2k^4|\D|^2+\lf(\xi_{\mbf k}^0\ri)^2},\\
\l_{\mbf k}^{a,\pm} &= \pm\sqrt{\fr{1}{2}k^4|\D|^2+\lf(\xi_{\mbf k}^a\ri)^2},\quad
a=\pm.
\end{align}
An illustration for the energy spectrum is presented in Fig.~\ref{FigdWPBdG}. The spectrum is fully gapped when the Fermi level is away from the band crossing points $\mu\neq0$. When the Fermi level lies exactly at the band crossing points $\mu=0$, a nodal point which separates the positive and negative bands emerges at $k=0$.

\subsubsection{Gap Function and Critical Temperature}

With the eigenvalues $\z_{\mbf k}^{a,\pm}$'s of the pairing representation $k^2\mca N_{00}^{22}$, the gap equation in this channel can be calculated from Eq.~(\ref{eq:gapeq})
\beeq
\beal
\fr{6}{V_2}
&=
\fr{1}{\mca V}\lf.\sum_{\mbf k}\ri.^0
\fr{k^4\tanh\fr{\b}{2}\sqrt{2k^4|\D|^2+(\xi_{\mbf k}^0)^2}}{\sqrt{2k^4|\D|^2+(\xi_{\mbf k}^0)^2}}
\\
&\quad
+\sum_{a=\pm}\fr{1}{\mca V}\lf.\sum_{\mbf k}\ri.^a
\fr{k^4\tanh\fr{\b}{2}\sqrt{k^4|\D|^2/2+(\xi_{\mbf k}^a)^2}}{4\sqrt{k^4|\D|^2/2+(\xi_{\mbf k}^a)^2}}.
\enal
\eneq
The flat band pairing contributes to the gap equation, which implies that the superconductivity can receive enhancement from the extremely large density of states $\nu^0(0)$ at the level of band crossing points $\pm P$. Therefore, the novel superconductivity with linear scaling in coupling strength $V_2$ can be supported in this channel. The superconducting properties in various regimes with different values of coupling strength $V_2$, chemical potential $\mu$, and temperature $T$ resemble those in the $s$-wave singlet channel (Sec.~\ref{sec:swp}).

In the regime where Fermi level is close to the band crossing points $|\mu|<\o_D$, the flat band pairing dominates the superconducting behavior. A suitable gap equation for this regime in the integral language Eq.~(\ref{eq:momspasumtoint}) is
\beeq
\label{eq:gapeqdwp}
\fr{6}{V_2}
=
\fr{1}{2\pi^2}\int_0^\L dkk^2
\fr{k^4\tanh\fr{\b}{2}\sqrt{2k^4|\D|^2+\mu^2}}{\sqrt{2k^4|\D|^2+\mu^2}}
.
\eneq
When the Fermi level lies exactly at the band crossing points $\mu=0$, the zero temperature gap function $\D(0)$ acquires exactly linear growth with increasing coupling strength
\beeq
\label{eq:lingapfundwp}
\lf|\D(0)\ri| = \fr{\L^5}{60\sqrt2\pi^2}V_2.
\eneq
Thermal fluctuation suppresses the superconductivity at finite temperature $T\neq0$. The critical temperature
\beeq
\label{eq:lincritemdwp}
T_c = \fr{\L^7}{168\pi^2}V_2
\eneq
marks the temperature where the superconductivity is fully eliminated, and grows linearly with increasing coupling strength $V_2$. The linear scalings Eq.~(\ref{eq:lingapfundwp}) and Eq.~(\ref{eq:lincritemdwp}) suggest that the $d$-wave quintet pairing channel supports novel superconductivity which is stronger and survives higher temperature than conventional BCS superconductivity. This pairing is suppressed by both finite doping and temperature in a manner analogous to the s-wave singlet pairing case discussed in Sec.~\ref{sec:swp}.


\section{Nonzero Band Curvature}
\label{sec:bandcurvature}

So far, we have kept only the leading order of $\mbf k\cdot\mbf p$ Hamiltonian in three band crossing materials. In this setup, a perfect flat band exists about each band crossing point $\pm P$. In actual band structures, however, the flat band may be perturbed by higher order corrections in the Hamiltonian. For the completeness of our studies of superconductivity in three band crossing materials, we examine the effect of such perturbations on the superconducting behavior of $J=0$ pairing states. For simplicity, we assume that the perturbations continue to respect spherical symmetry.

We add a quadratic term to the $\mbf k\cdot\mbf p$ Hamiltonian at $P$ Eq.~(\ref{EqHP}) as a perturbation. With the preservation of full spherical symmetry about three band crossing point $P$, the quadratic perturbation term takes the form
\beeq
\d H_P(\mbf k) = \fr{\sqrt2}{2m}\lf|\mbf k-\mbf k_0\ri|^2\mca N_{00}^{22}.
\eneq
Since the irreducible representation $\mca N_{00}^{22}$ is quadratic in spin operators $S^i$'s, it is even under TR transformation Eq.~(\ref{EqTRST})
\beeq
\mca T\mca N_{00}^{22}\mca T^{-1} = \mca N_{00}^{22}.
\eneq
To preserve TR symmetry Eq.~(\ref{eq:TRS}) of the whole system, the Hamiltonians at $\pm P$ are given by
\beeq
H_{\pm P}(\pm\mbf k)
= v\lf[\pm\lf(\mbf k-\mbf k_0\ri)\ri]\cdot\mbf S
+\fr{\sqrt2}{2m}\lf|\pm\lf(\mbf k-\mbf k_0\ri)\ri|^2\mca N_{00}^{22}.
\eneq
Notice the same sign for the quadratic terms in $H_\pm(\mbf k)$.

With the addition of quadratic perturbation term, the noninteracting Hamiltonian $\mca H_0(\mbf k)$ Eq.~(\ref{eq:nonintham}) in the BdG Hamiltonian Eq.~(\ref{eq:bdgham}) becomes
\beeq
\mca H_0(\mbf k)
= v\mbf k\cdot\mbf S+\sqrt2\fr{k^2\mca N_{00}^{22}}{2m}-\mu.
\eneq
The eigenvalues are
\beeq
\xi_{\mbf k}^0 = -\fr{k^2}{m}-\mu,\quad
\xi_{\mbf k}^\pm = \pm vk+\fr{k^2}{2m}-\mu.
\eneq
 Notice that the noninteracting Hamiltonian $\mca H_0(\mbf k)$ still commutes with the pairing representations in all three $J=0$ irreducible pairing channels, implying the applicability of our setup in Sec.~\ref{sec:scspinonefermion}.

We investigate the effect of finite band curvature on the superconductivity in the regime $|\mu|<\o_D$, where the Fermi level is close to the band crossing points. The linearly dispersing bands do not change significantly in this regime. Therefore, the $p$-wave triplet $(1,1,0)$ pairing state, which receives only contributions from linearly dispersing bands, still exhibits similar superconducting behavior to the nonperturbed theory. The $s$-wave singlet $(0,0,0)$ and $d$-wave quintet $(2,2,0)$ pairing states, however, are affected. Finite band curvature leads to a significant change in the density of states of flat band $\nu^0(\ve)$, and therefore brings about different superconducting behavior from the nonperturbed theory.

\begin{figure}[t]
\includegraphics[scale=1]{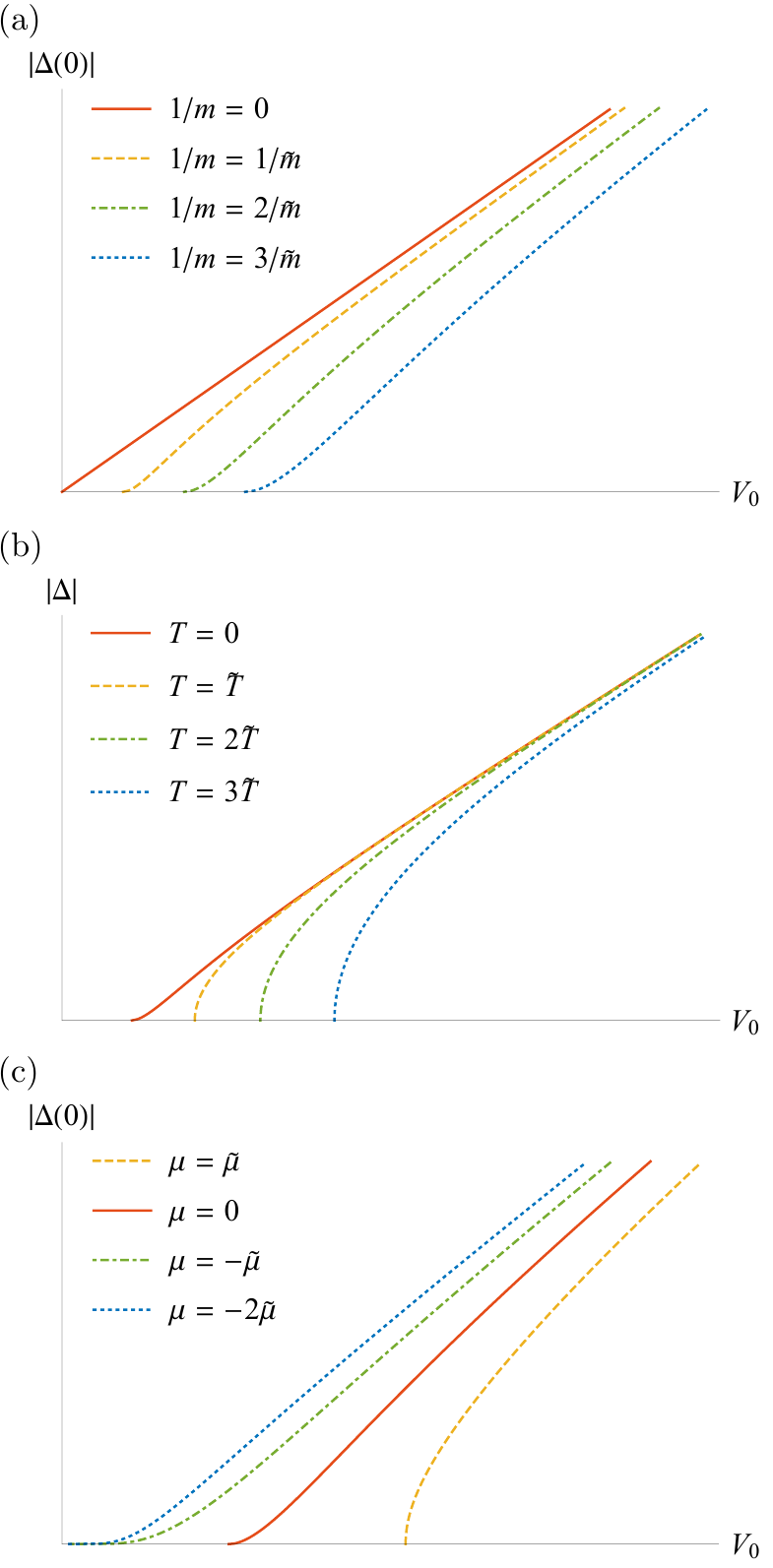}
\caption{\label{fig:bcswpgf} Numerically obtained gap function in the $s$-wave singlet pairing channel with band curvature. Here $\til m$, $\til\mu$, and $\til T$ are positive constants. (a) At $\mu=0$ and $T=0$, finite band curvature $1/m$ shifts the threshold coupling strength $V_0^\mrm{th}$ linearly. (b) Thermal fluctuations suppress the superconductivity. Here $\mu=0$ and $m=\til m$. (c) While positive $\mu$ suppress the superconductivity, negative $\mu$ provides enhancement in weak coupling regime. Here $T=0$ and $m=\til m$.}
\end{figure}

We first study the effect of perturbation on $s$-wave singlet pairing state. With the observation of dominating flat band contribution, the gap equation Eq.~(\ref{eq:gapeqswpmomspasum}) reads
\beeq
\fr{6}{V_0}
= \fr{1}{\mca V}\lf.\sum_{\mbf k}\ri.^0\fr{\tanh\fr{\b}{2}\sqrt{|\D|^2+(-k^2/m-\mu)^2}}{2\sqrt{|\D|^2+(-k^2/m-\mu)^2}}.
\eneq
As in the nonpeturbed theory, we transform the momentum space summation into a momentum space integral within the nearly flat band spherical domain $k<\L$ Eq.~(\ref{eq:momspasumtoint}). The integral version of gap equation takes the form
\beeq
\fr{6}{V_0}
= \fr{1}{2\pi^2}\int_0^\L dkk^2\fr{\tanh\fr{\b}{2}\sqrt{|\D|^2+(-k^2/m-\mu)^2}}{2\sqrt{|\D|^2+(-k^2/m-\mu)^2}}.
\eneq
A significant difference from the nonperturbed theory is the existence of a finite threshold coupling strength $V_0=V_0^\mrm{th}$ at $T=0$ when the Fermi level lies at the band crossing points $\mu=0$ [Fig.~\ref{fig:bcswpgf}(a)]
\beeq
V_0^\mrm{th} = \fr{24\pi^2}{\L}\fr{1}{m}.
\eneq
The absence of superconductivity in weak coupling regime results from the vanishing density of states $\nu^0(\ve)=0$ induced by the finite band curvature. However, since the band curvature is perturbative (i.e. $1/m$ is small), the interaction required for turning on superconductivity is small. As the coupling strength $V_0$ increases beyond the threshold value $V_0=V_0^\mrm{th}$, the zero temperature gap function $\D(0)$ recovers the linear scaling Eq.~(\ref{eq:lingapfunswp}) in strong coupling regime. Notice the linear dependence of threshold coupling strength $V_0^\mrm{th}$ on band curvature $1/m$. The nonperturbed theory $V_0^\mrm{th}=0$ is recovered in the zero curvature limit $1/m\rar0$. These results are illustrated in Fig.~\ref{fig:bcswpgf}(a).

\begin{figure}[t]
\includegraphics[scale=0.5]{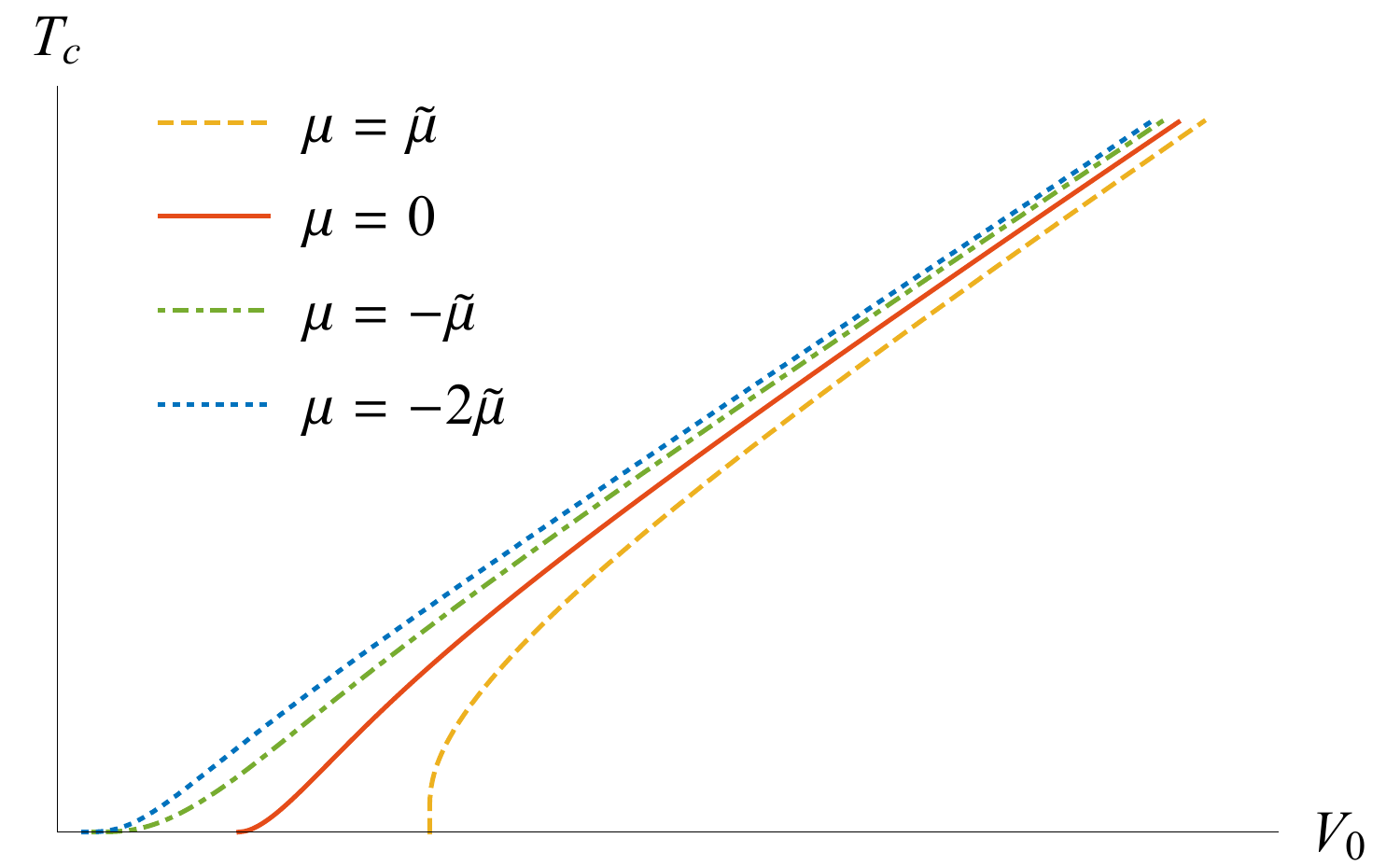}
\caption{\label{fig:bcswptc} Critical temperature as a function of coupling strength in the $s$-wave singlet $(0,0,0)$ pairing channel with band curvature. We choose the band curvature $1/m=1/\til m$ for some positive constant $\til m$. Another positive constant $\til\mu$ is chosen for plotting. It can be observed that the critical temperature increases when the chemical potential goes down from zero.}
\end{figure}

We also investigate the effect of finite temperature $T$ and finite chemical potential $\mu$ on the superconductivity. The suppression of superconductivity by thermal fluctuations is similar to the nonperturbed theory [Fig.~\ref{fig:bcswpgf}(b)]. Raising the Fermi level from band crossing points $\mu>0$ makes it farther from the middle band, which leads to the suppression, as well [Fig.~\ref{fig:bcswpgf}(c)]. However, a different behavior arises when the Fermi level is lowered from the band crossing points $\mu<0$. The Fermi level crosses the middle band and creates a finite Fermi surface. Due to the finite density of states $\nu^0(\mu)$, the conventional superconductivity is supported in the weak coupling regime. The zero temperature gap function $\D(0)$ starts to grow once the interaction is turned on, and the exponentially small scaling is observed. As $\mu$ goes down, the growing density of states implies the strengthening of superconductivity, with the zero temperature gap function and critical temperature (Fig.~\ref{fig:bcswptc}) increasing. The linear scaling in coupling strength is recovered as the interaction becomes stronger.

The behavior of $d$-wave quintet pairing state with finite band curvature is similar to the $s$-wave singlet pairing state. We note that with the gap equation
\beeq
\fr{6}{V_2}
= \fr{1}{2\pi^2}\int_0^\L dkk^2\fr{k^4\tanh\fr{\b}{2}\sqrt{2k^4|\D|^2+(-k^2/m-\mu)^2}}{\sqrt{2k^4|\D|^2+(-k^2/m-\mu)^2}},
\eneq
the threshold coupling strength for zero temperature gap function $\D(0)$ at $\mu=0$ can be determined
\beeq
V_2 = \fr{60\pi^2}{\L^5}\fr{1}{m}.
\eneq
The linear dependence of threshold coupling strength $V_2^\mrm{th}$ on band curvature $1/m$ still holds in this channel.

\section{Discussion and Conclusion}
\label{SecDisConc}

In this work, we have studied the pairing problem of spin one fermions, which arises near three band crossing points in certain topological band structures. We have identified the minimal BCS Hamiltonian, decomposed it into spin-orbit coupled irreducible pairing channels labelled by $(L,S,J)$, and have obtained the gap equation in each channel. We have solved these gap equations for the three channels with zero total angular momentum $(L,S,J)=(0,0,0)$, $(1,1,0)$, and $(2,2,0)$ with the Fermi level at the band crossing points. Discussions on the effects of finite temperature and finite displacement of Fermi level from band crossing points are also presented. Interesting results show up when the Fermi level is close enough to the band crossing points. Otherwise one may simply project the weak coupling pairing problem onto a single band and eliminate the novel features of the spin one problem. We have found that 
in the $s$-wave spin singlet and $d$-wave spin quintet pairing channels, pairing is enormously enhanced. In particular, when the Fermi level lies exactly at the band crossing points, the critical temperature is {\it linear} in interaction strength. This enormous enhancement arises because of a flat band with a concomitant large density of states. When the Fermi level is shifted slightly away from the band crossing points, the linear characteristic can still show up with strong enough interaction. Meanwhile, superconductivity exhibits features of conventional BCS theory in the $p$-wave spin triplet pairing channel, since this pairing channel is not able to take advantage of the flat band. In addition to the perfect flat band setup, the effect of finite band curvature on superconducting behavior is also examined, and the modifications of weak coupling theories in $s$-wave and $d$-wave pairing states are discovered. These results suggest that three band crossings may represent good platforms for exotic superconductivity (e.g. spin quintet) with enormously enhanced energy scales. 

This work thus opens a new direction in the study of superconductivity in topological band structures. Much, however, remains to be explored. For example, the current work focuses on single-component pairing channels. In general, the three band crossing materials can also support multi-component pairing channels, where the total angular momentum $J$'s are nonzero and the gap functions are multi-component vectors \cite{Venderbos, Boettcher}. The studies of multi-component pairing channels requires  Landau-Ginzburg analysis, and would be an interesting topic for future work. One can also consider the superconductivity with finite momentum pairing, which is known as the Fulde-Ferrell-Larkin-Ovchinnikov (FFLO) state \cite{Fulde1964PR, larkin1964}. We did not consider FFLO pairing in this work. Since the noninteracting Hamiltonians at $\pm P$ are identical at leading order in the $\mathbf{k} \cdot \mathbf{p}$ expansion, the zero momentum and FFLO pairings will be degenerate at this order. The degeneracy will be lifted when higher order terms in the $\mathbf{k} \cdot \mathbf{p}$ expansion are retained, presumably in favor of the zero momentum pairing states. However a careful investigation of the possibilities of FFLO pairing in such materials would also be an interesting direction for future work. 

Additionally, the analysis in this manuscript assumed full spherical symmetry, such that the angular momenta constituted good quantum numbers. While this symmetry is indeed present in the low energy Hamiltonian, the full problem will have only the lower symmetry of the lattice space group. The decomposition of the pairing problem into irreps of the lattice space group, and the solution of the resulting gap equations, will differ in details from the analysis presented herein, and is also left as a problem for future work. Furthermore, the present work has focused on mean field analysis in clean (disorder free) systems. Going beyond mean field, or including the effects of disorder on superconductivity, would both be interesting challenges in their own right, as would an analysis of correlated states besides superconductivity. These problems too are left to future work. Finally, the large density of states of the flat band could also enhance alternative instabilities e.g. to density waves. The investigation of the competition between superconductivity and other orders would also be an interesting problem for future work. 

The most important conclusion of this work, however, is the following: three band crossings represent good platforms for realizing superconductivity with enormously enhanced energy scales. We eagerly anticipate experimental searches for superconductivity in this setting. 


\section{Acknowledgements}
We acknowledge valuable conversations with Yang-Zhi Chou. We also thank Yang-Zhi Chou, J\"orn Venderbos, and Yuxuan Wang for feedback on the manuscript. This research was sponsored by the Army Research Office and was accomplished under Grant Number W911NF-17-1-0482. The views and conclusions contained in this document are those of the authors and should not be interpreted as representing the official policies, either expressed or implied, of the Army Research Office or the U.S. Government. The U.S. Government is authorized to reproduce and distribute reprints for Government purposes notwithstanding any copyright notation herein.

\section*{Appendix: Superconductivity from Pairing of Band Crossing Points with Different Signs}

\begin{figure}[t]
\includegraphics[scale=1]{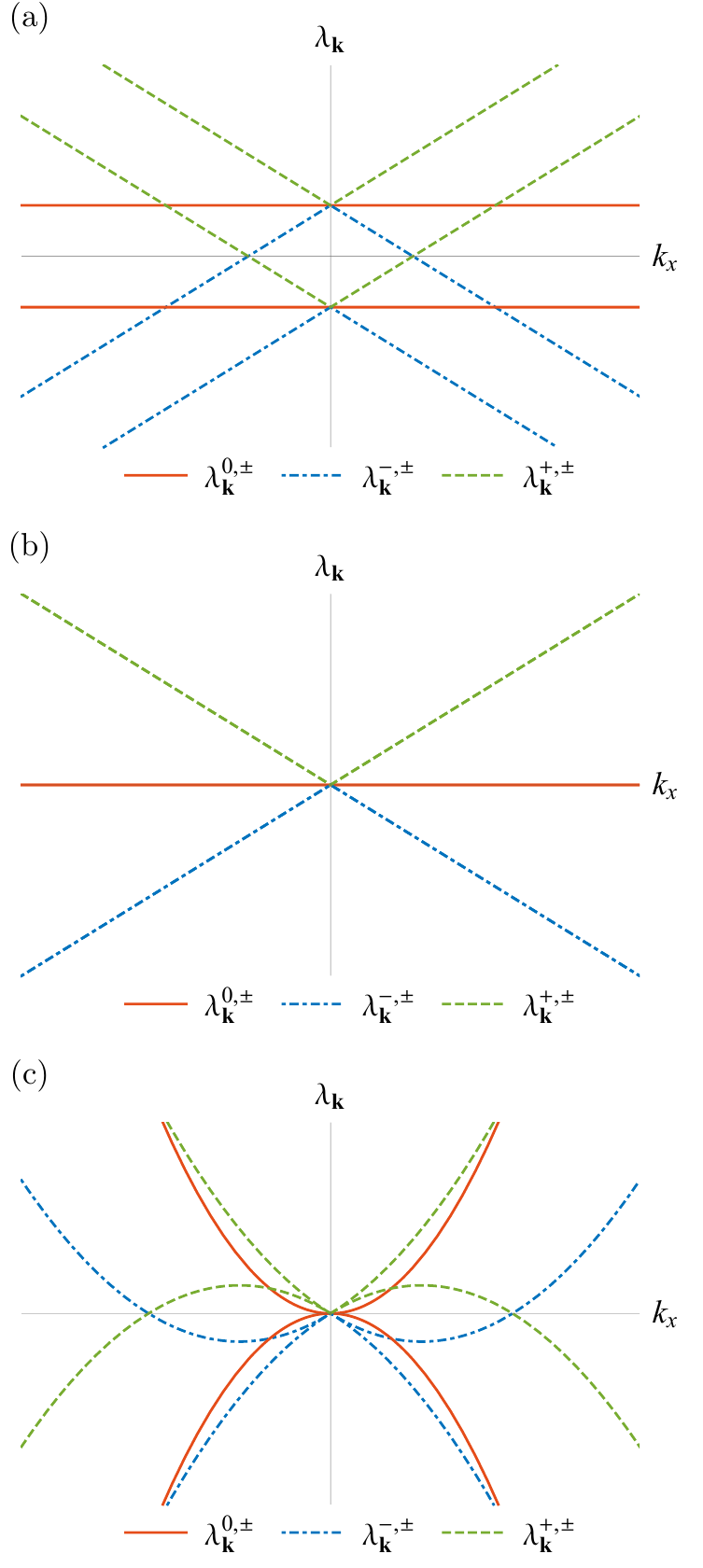}
\caption{\label{fig:osbdg} BdG quasiparticle energy spectrum at $\mu=0$ for (a) $s$-wave (b) $p$-wave and (c) $d$-wave $J=0$ pairing channels, in the case when the noninteracting Hamiltonians at $\pm P$ take opposite sign. Nodal shell exists in $s$-wave and $d$-wave pairing channels with radius $|\D|/v$ and $\sqrt2v/|\D|$, respectively. The latter vanishes in weak coupling regime. In $p$-wave pairing channel, superconductivity is absent and $\D=0$. Each band is two fold degenerate due to the overlap of bands about $\pm P$.}
\end{figure}

We have conducted a comprehensive investigation of the potential superconductivities in the TR symmetric three band crossing materials, where the $\mbf k\cdot\mbf S$ Hamiltonians at band crossing points $\pm P$ take same sign. As a complementary study, we briefly examine the superconductivity in the systems where the Hamiltonians at the band crossing points have opposite signs $H_{\pm P}(\pm\mbf k) = \pm(\pm\mbf k)\cdot\mbf S$. This type of Hamiltonian may describe, for example, three band crossing materials with inversion symmetry. 

For the $J=0$ irreducible pairing channels, the whole BdG Hamiltonian can be decomposed into the BdG Hamiltonians $\mca H_\mrm{BdG}^{(L,L,0),a}(\mbf k)$'s of different $\ket{u_{\mbf k}^a}$-sections Eq.~(\ref{eq:bdgblkdiag}). The single band BdG Hamiltonian $\mca H_\mrm{BdG}^{(L,L,0),a}(\mbf k)$ is similar to that in the TR symmetric systems Eq.~(\ref{eq:bdghamnoninteigmode}). Due to the changed sign in $H_{-P}(-\mbf k)$, the sign of noninteracting band energy $\ve_{\mbf k}^a$ in the $(2,2)$-element of $\mca H_\mrm{BdG}^{(L,L,0),a}(\mbf k)$ is flipped
\beeq
\mca H_\mrm{BdG}^{(L,L,0),a}(\mbf k) =
\lf(\bear{cc}
\ve_{\mbf k}^a-\mu & -\D\z_{\mbf k}^a\\
-\bar\D\z_{\mbf k}^a&\ve_{\mbf k}^a+\mu
\enar\ri),\quad
a=0,\pm.
\eneq
The BdG quasiparticle energies $\l_{\mbf k}^a$'s become
\beeq
\l_{\mbf k}^{a,\pm}
= \ve_{\mbf k}^a\pm\sqrt{|\D|^2(\z_{\mbf k}^a)^2+\mu^2},\quad
a=0,\pm,
\eneq
indicating that the Cooper pairing modifies the noninteracting energies $\ve_{\mbf k}^a$'s by a momentum dependent shift (Fig.~\ref{fig:osbdg}). The resulting BdG quasiparticle energy spectrums can contain some nontrivial nodal structures in the bulk, which are different from the spectrums in TR symmetric systems.

Substituting the BdG quasiparticle energies into the gap equation Eq.~(\ref{eq:gapeqf}) yields
\beeq
\beal
\fr{6\bar\D}{V_L}
= -&\sum_{a=0,\pm}
\fr{\bar\D}{\mca V}\lf.\sum_{\mbf k}\ri.^a
\\
&~
\sum_{b=\pm}
\fr{b(\z_{\mbf k}^a)^2 f(\ve_{\mbf k}^a+b\sqrt{|\D|^2[\z_{\mbf k}^a]^2+\mu^2})}{2\sqrt{|\D|^2(\z_{\mbf k}^a)^2+\mu^2}}
.
\enal
\eneq
With the exact form of Fermi function $f(z)$ Eq.~(\ref{eq:fermifunction}), the gap equation for nontrivial gap function $\D\neq0$ is
\beeq
\beal
\fr{6}{V_L}
= \sum_{a=0,\pm}
&\fr{1}{\mca V}\lf.\sum_{\mbf k}\ri.^a
\fr{(\z_{\mbf k}^a)^2}{2\sqrt{|\D|^2(\z_{\mbf k}^a)^2+\mu^2}}
\\
&
\times\fr{\sinh\b\sqrt{|\D|^2(\z_{\mbf k}^a)^2+\mu^2}}{\cosh\b\ve_{\mbf k}^a+\cosh\b\sqrt{|\D|^2(\z_{\mbf k}^a)^2+\mu^2}}
.
\enal
\eneq
The potential superconductivity in each irreducible pairing channel can be probed by examining the solutions of this gap equation.

We focus on the regime $|\mu|<\o_D$, where the Fermi level is close to the band crossing points and novel superconductivities appear in the TR symmetric systems (Sec.~\ref{sec:scspinonefermion}). For the $s$-wave singlet $(0,0,0)$ and $d$-wave quintet $(2,2,0)$ pairing channels, the flat band pairing dominates the pairing state. Neglecting the contribution from linearly dispersing bands, the gap equations are exactly the same as Eq.~(\ref{eq:gapeqswp}) and Eq.~(\ref{eq:gapeqdwp}) in the TR symmetric systems. Therefore, the novel superconductivity with linear scaling in coupling strength shows up in this regime. The superconducting behavior is the same as in the TR symmetric systems, except for the difference in infinitesimal corrections from linearly dispersing bands.

For the $p$-wave triplet $(1,1,0)$ pairing channel, the gap equation takes the form
\beeq
\beal
\fr{6}{V_1}
= \sum_{a=\pm}
&\fr{1}{\mca V}\lf.\sum_{\mbf k}\ri.^a
\fr{3k^2}{4\sqrt{3k^2|\D|^2/2+\mu^2}}
\\
&
\times\fr{\sinh\b\sqrt{3k^2|\D|^2/2+\mu^2}}{\cosh\b vk+\cosh\b\sqrt{3k^2|\D|^2/2+\mu^2}}
.
\enal
\eneq
The flat band does not involve in the Cooper pairing, implying the absence of novel superconductivity which appears in the other two $J=0$ pairing channels. As an examination, we inspect the behavior of pairing state at $\mu=0$ in the weak coupling regime. In the zero temperature limit $\b\rar\infty$, the second fractional number in the gap equation reduces to a step function
\beeq
\beal
&\lim_{\b\rar\infty}\fr{\sinh\b\sqrt{3k^2|\D|^2/2+\mu^2}}{\cosh\b vk+\cosh\b\sqrt{3k^2|\D|^2/2+\mu^2}}
\\
&= \t\lf(\lf[3|\D|^2/2-v^2\ri]k^2+\mu^2\ri).
\enal
\eneq
Notice that the gap function $\D$ should be infinitesimal in the weak coupling regime. When $\mu=0$, the step function always vanishes, and the gap equation does not support nontrivial solution. Therefore, superconductivity does not exist in weak coupling regime.

Note that in the inversion but not TR symmetric case, when the noninteracting Hamiltonians at $\pm P$ have opposite sign, zero momentum pairing states are generally gapless, containing `nodal shells.' In contrast, FFLO pairing states, involving only a single `valley', will presumably be fully gapped (similar to the states discussed in the main text). As such, FFLO pairing states may be expected to maximize condensation energy better than zero momentum pairing states. We therefore speculate that inversion but not TR symmetric three band crossings (if such can be realized) may be a good platform for stabilizing FFLO pairing with enhanced energy scales. A detailed discussion of this issue is left to future work.

\bibliography{Reference}

\begin{thebibliography}{23}%
\makeatletter
\providecommand \@ifxundefined [1]{%
 \@ifx{#1\undefined}
}%
\providecommand \@ifnum [1]{%
 \ifnum #1\expandafter \@firstoftwo
 \else \expandafter \@secondoftwo
 \fi
}%
\providecommand \@ifx [1]{%
 \ifx #1\expandafter \@firstoftwo
 \else \expandafter \@secondoftwo
 \fi
}%
\providecommand \natexlab [1]{#1}%
\providecommand \enquote  [1]{``#1''}%
\providecommand \bibnamefont  [1]{#1}%
\providecommand \bibfnamefont [1]{#1}%
\providecommand \citenamefont [1]{#1}%
\providecommand \href@noop [0]{\@secondoftwo}%
\providecommand \href [0]{\begingroup \@sanitize@url \@href}%
\providecommand \@href[1]{\@@startlink{#1}\@@href}%
\providecommand \@@href[1]{\endgroup#1\@@endlink}%
\providecommand \@sanitize@url [0]{\catcode `\\12\catcode `\$12\catcode
  `\&12\catcode `\#12\catcode `\^12\catcode `\_12\catcode `\%12\relax}%
\providecommand \@@startlink[1]{}%
\providecommand \@@endlink[0]{}%
\providecommand \url  [0]{\begingroup\@sanitize@url \@url }%
\providecommand \@url [1]{\endgroup\@href {#1}{\urlprefix }}%
\providecommand \urlprefix  [0]{URL }%
\providecommand \Eprint [0]{\href }%
\providecommand \doibase [0]{http://dx.doi.org/}%
\providecommand \selectlanguage [0]{\@gobble}%
\providecommand \bibinfo  [0]{\@secondoftwo}%
\providecommand \bibfield  [0]{\@secondoftwo}%
\providecommand \translation [1]{[#1]}%
\providecommand \BibitemOpen [0]{}%
\providecommand \bibitemStop [0]{}%
\providecommand \bibitemNoStop [0]{.\EOS\space}%
\providecommand \EOS [0]{\spacefactor3000\relax}%
\providecommand \BibitemShut  [1]{\csname bibitem#1\endcsname}%
\let\auto@bib@innerbib\@empty
\bibitem [{\citenamefont {Armitage}\ \emph {et~al.}(2018)\citenamefont
  {Armitage}, \citenamefont {Mele},\ and\ \citenamefont
  {Vishwanath}}]{Armitege2018RMP}%
  \BibitemOpen
  \bibfield  {author} {\bibinfo {author} {\bibfnamefont {N.~P.}\ \bibnamefont
  {Armitage}}, \bibinfo {author} {\bibfnamefont {E.~J.}\ \bibnamefont {Mele}},
  \ and\ \bibinfo {author} {\bibfnamefont {A.}~\bibnamefont {Vishwanath}},\
  }\href {\doibase 10.1103/RevModPhys.90.015001} {\bibfield  {journal}
  {\bibinfo  {journal} {Rev. Mod. Phys.}\ }\textbf {\bibinfo {volume} {90}},\
  \bibinfo {pages} {015001} (\bibinfo {year} {2018})}\BibitemShut {NoStop}%
\bibitem [{\citenamefont {Bradlyn}\ \emph {et~al.}(2016)\citenamefont
  {Bradlyn}, \citenamefont {Cano}, \citenamefont {Wang}, \citenamefont
  {Vergniory}, \citenamefont {Felser}, \citenamefont {Cava},\ and\
  \citenamefont {Bernevig}}]{Bradlyn2016Sci}%
  \BibitemOpen
  \bibfield  {author} {\bibinfo {author} {\bibfnamefont {B.}~\bibnamefont
  {Bradlyn}}, \bibinfo {author} {\bibfnamefont {J.}~\bibnamefont {Cano}},
  \bibinfo {author} {\bibfnamefont {Z.}~\bibnamefont {Wang}}, \bibinfo {author}
  {\bibfnamefont {M.~G.}\ \bibnamefont {Vergniory}}, \bibinfo {author}
  {\bibfnamefont {C.}~\bibnamefont {Felser}}, \bibinfo {author} {\bibfnamefont
  {R.~J.}\ \bibnamefont {Cava}}, \ and\ \bibinfo {author} {\bibfnamefont
  {B.~A.}\ \bibnamefont {Bernevig}},\ }\href {\doibase 10.1126/science.aaf5037}
  {\bibfield  {journal} {\bibinfo  {journal} {Science}\ }\textbf {\bibinfo
  {volume} {353}},\ \bibinfo {pages} {aaf5037} (\bibinfo {year}
  {2016})}\BibitemShut {NoStop}%
\bibitem [{\citenamefont {Fu}\ and\ \citenamefont {Kane}(2008)}]{FuKane}%
  \BibitemOpen
  \bibfield  {author} {\bibinfo {author} {\bibfnamefont {L.}~\bibnamefont
  {Fu}}\ and\ \bibinfo {author} {\bibfnamefont {C.~L.}\ \bibnamefont {Kane}},\
  }\href {\doibase 10.1103/PhysRevLett.100.096407} {\bibfield  {journal}
  {\bibinfo  {journal} {Phys. Rev. Lett.}\ }\textbf {\bibinfo {volume} {100}},\
  \bibinfo {pages} {096407} (\bibinfo {year} {2008})}\BibitemShut {NoStop}%
\bibitem [{\citenamefont {Li}\ and\ \citenamefont {Haldane}(2018)}]{Li2018PRL}%
  \BibitemOpen
  \bibfield  {author} {\bibinfo {author} {\bibfnamefont {Y.}~\bibnamefont
  {Li}}\ and\ \bibinfo {author} {\bibfnamefont {F.~D.~M.}\ \bibnamefont
  {Haldane}},\ }\href {\doibase 10.1103/PhysRevLett.120.067003} {\bibfield
  {journal} {\bibinfo  {journal} {Phys. Rev. Lett.}\ }\textbf {\bibinfo
  {volume} {120}},\ \bibinfo {pages} {067003} (\bibinfo {year}
  {2018})}\BibitemShut {NoStop}%
\bibitem [{\citenamefont {Yang}\ \emph {et~al.}(2016)\citenamefont {Yang},
  \citenamefont {Li},\ and\ \citenamefont {Wu}}]{congjun}%
  \BibitemOpen
  \bibfield  {author} {\bibinfo {author} {\bibfnamefont {W.}~\bibnamefont
  {Yang}}, \bibinfo {author} {\bibfnamefont {Y.}~\bibnamefont {Li}}, \ and\
  \bibinfo {author} {\bibfnamefont {C.}~\bibnamefont {Wu}},\ }\href {\doibase
  10.1103/PhysRevLett.117.075301} {\bibfield  {journal} {\bibinfo  {journal}
  {Phys. Rev. Lett.}\ }\textbf {\bibinfo {volume} {117}},\ \bibinfo {pages}
  {075301} (\bibinfo {year} {2016})}\BibitemShut {NoStop}%
\bibitem [{\citenamefont {Brydon}\ \emph {et~al.}(2016)\citenamefont {Brydon},
  \citenamefont {Wang}, \citenamefont {Weinert},\ and\ \citenamefont
  {Agterberg}}]{agterberg}%
  \BibitemOpen
  \bibfield  {author} {\bibinfo {author} {\bibfnamefont {P.~M.~R.}\
  \bibnamefont {Brydon}}, \bibinfo {author} {\bibfnamefont {L.}~\bibnamefont
  {Wang}}, \bibinfo {author} {\bibfnamefont {M.}~\bibnamefont {Weinert}}, \
  and\ \bibinfo {author} {\bibfnamefont {D.~F.}\ \bibnamefont {Agterberg}},\
  }\href {\doibase 10.1103/PhysRevLett.116.177001} {\bibfield  {journal}
  {\bibinfo  {journal} {Phys. Rev. Lett.}\ }\textbf {\bibinfo {volume} {116}},\
  \bibinfo {pages} {177001} (\bibinfo {year} {2016})}\BibitemShut {NoStop}%
\bibitem [{\citenamefont {Kim}\ \emph {et~al.}(2018)\citenamefont {Kim},
  \citenamefont {Wang}, \citenamefont {Nakajima}, \citenamefont {Hu},
  \citenamefont {Ziemak}, \citenamefont {Syers}, \citenamefont {Wang},
  \citenamefont {Hodovanets}, \citenamefont {Denlinger}, \citenamefont
  {Brydon}, \citenamefont {Agterberg}, \citenamefont {Tanatar}, \citenamefont
  {Prozorov},\ and\ \citenamefont {Paglione}}]{paglione}%
  \BibitemOpen
  \bibfield  {author} {\bibinfo {author} {\bibfnamefont {H.}~\bibnamefont
  {Kim}}, \bibinfo {author} {\bibfnamefont {K.}~\bibnamefont {Wang}}, \bibinfo
  {author} {\bibfnamefont {Y.}~\bibnamefont {Nakajima}}, \bibinfo {author}
  {\bibfnamefont {R.}~\bibnamefont {Hu}}, \bibinfo {author} {\bibfnamefont
  {S.}~\bibnamefont {Ziemak}}, \bibinfo {author} {\bibfnamefont
  {P.}~\bibnamefont {Syers}}, \bibinfo {author} {\bibfnamefont
  {L.}~\bibnamefont {Wang}}, \bibinfo {author} {\bibfnamefont {H.}~\bibnamefont
  {Hodovanets}}, \bibinfo {author} {\bibfnamefont {J.~D.}\ \bibnamefont
  {Denlinger}}, \bibinfo {author} {\bibfnamefont {P.~M.~R.}\ \bibnamefont
  {Brydon}}, \bibinfo {author} {\bibfnamefont {D.~F.}\ \bibnamefont
  {Agterberg}}, \bibinfo {author} {\bibfnamefont {M.~A.}\ \bibnamefont
  {Tanatar}}, \bibinfo {author} {\bibfnamefont {R.}~\bibnamefont {Prozorov}}, \
  and\ \bibinfo {author} {\bibfnamefont {J.}~\bibnamefont {Paglione}},\ }\href
  {\doibase 10.1126/sciadv.aao4513} {\bibfield  {journal} {\bibinfo  {journal}
  {Sci. Adv.}\ }\textbf {\bibinfo {volume} {4}},\ \bibinfo {pages} {aao4513}
  (\bibinfo {year} {2018})}\BibitemShut {NoStop}%
\bibitem [{\citenamefont {Boettcher}\ and\ \citenamefont
  {Herbut}(2016)}]{Herbut}%
  \BibitemOpen
  \bibfield  {author} {\bibinfo {author} {\bibfnamefont {I.}~\bibnamefont
  {Boettcher}}\ and\ \bibinfo {author} {\bibfnamefont {I.~F.}\ \bibnamefont
  {Herbut}},\ }\href {\doibase 10.1103/PhysRevB.93.205138} {\bibfield
  {journal} {\bibinfo  {journal} {Phys. Rev. B}\ }\textbf {\bibinfo {volume}
  {93}},\ \bibinfo {pages} {205138} (\bibinfo {year} {2016})}\BibitemShut
  {NoStop}%
\bibitem [{\citenamefont {Savary}\ \emph {et~al.}(2017)\citenamefont {Savary},
  \citenamefont {Ruhman}, \citenamefont {Venderbos}, \citenamefont {Fu},\ and\
  \citenamefont {Lee}}]{Savary2017PRB}%
  \BibitemOpen
  \bibfield  {author} {\bibinfo {author} {\bibfnamefont {L.}~\bibnamefont
  {Savary}}, \bibinfo {author} {\bibfnamefont {J.}~\bibnamefont {Ruhman}},
  \bibinfo {author} {\bibfnamefont {J.~W.~F.}\ \bibnamefont {Venderbos}},
  \bibinfo {author} {\bibfnamefont {L.}~\bibnamefont {Fu}}, \ and\ \bibinfo
  {author} {\bibfnamefont {P.~A.}\ \bibnamefont {Lee}},\ }\href {\doibase
  10.1103/PhysRevB.96.214514} {\bibfield  {journal} {\bibinfo  {journal} {Phys.
  Rev. B}\ }\textbf {\bibinfo {volume} {96}},\ \bibinfo {pages} {214514}
  (\bibinfo {year} {2017})}\BibitemShut {NoStop}%
\bibitem [{\citenamefont {Boettcher}\ and\ \citenamefont
  {Herbut}(2018)}]{Boettcher}%
  \BibitemOpen
  \bibfield  {author} {\bibinfo {author} {\bibfnamefont {I.}~\bibnamefont
  {Boettcher}}\ and\ \bibinfo {author} {\bibfnamefont {I.~F.}\ \bibnamefont
  {Herbut}},\ }\href {\doibase 10.1103/PhysRevLett.120.057002} {\bibfield
  {journal} {\bibinfo  {journal} {Phys. Rev. Lett.}\ }\textbf {\bibinfo
  {volume} {120}},\ \bibinfo {pages} {057002} (\bibinfo {year}
  {2018})}\BibitemShut {NoStop}%
\bibitem [{\citenamefont {Venderbos}\ \emph {et~al.}(2018)\citenamefont
  {Venderbos}, \citenamefont {Savary}, \citenamefont {Ruhman}, \citenamefont
  {Lee},\ and\ \citenamefont {Fu}}]{Venderbos}%
  \BibitemOpen
  \bibfield  {author} {\bibinfo {author} {\bibfnamefont {J.~W.~F.}\
  \bibnamefont {Venderbos}}, \bibinfo {author} {\bibfnamefont {L.}~\bibnamefont
  {Savary}}, \bibinfo {author} {\bibfnamefont {J.}~\bibnamefont {Ruhman}},
  \bibinfo {author} {\bibfnamefont {P.~A.}\ \bibnamefont {Lee}}, \ and\
  \bibinfo {author} {\bibfnamefont {L.}~\bibnamefont {Fu}},\ }\href {\doibase
  10.1103/PhysRevX.8.011029} {\bibfield  {journal} {\bibinfo  {journal} {Phys.
  Rev. X}\ }\textbf {\bibinfo {volume} {8}},\ \bibinfo {pages} {011029}
  (\bibinfo {year} {2018})}\BibitemShut {NoStop}%
\bibitem [{\citenamefont {Ghorashi}\ \emph {et~al.}(2017)\citenamefont
  {Ghorashi}, \citenamefont {Davis},\ and\ \citenamefont
  {Foster}}]{Ghorashi2017PRB}%
  \BibitemOpen
  \bibfield  {author} {\bibinfo {author} {\bibfnamefont {S.~A.~A.}\
  \bibnamefont {Ghorashi}}, \bibinfo {author} {\bibfnamefont {S.}~\bibnamefont
  {Davis}}, \ and\ \bibinfo {author} {\bibfnamefont {M.~S.}\ \bibnamefont
  {Foster}},\ }\href {\doibase 10.1103/PhysRevB.95.144503} {\bibfield
  {journal} {\bibinfo  {journal} {Phys. Rev. B}\ }\textbf {\bibinfo {volume}
  {95}},\ \bibinfo {pages} {144503} (\bibinfo {year} {2017})}\BibitemShut
  {NoStop}%
\bibitem [{\citenamefont {Roy}\ \emph {et~al.}(2017)\citenamefont {Roy},
  \citenamefont {Ghorashi}, \citenamefont {Foster},\ and\ \citenamefont
  {Nevidomskyy}}]{Roy2017Arxiv}%
  \BibitemOpen
  \bibfield  {author} {\bibinfo {author} {\bibfnamefont {B.}~\bibnamefont
  {Roy}}, \bibinfo {author} {\bibfnamefont {S.~A.~A.}\ \bibnamefont
  {Ghorashi}}, \bibinfo {author} {\bibfnamefont {M.~S.}\ \bibnamefont
  {Foster}}, \ and\ \bibinfo {author} {\bibfnamefont {A.~H.}\ \bibnamefont
  {Nevidomskyy}},\ }\href@noop {} {\  (\bibinfo {year} {2017})},\ \Eprint
  {http://arxiv.org/abs/1708.07825} {arXiv:1708.07825 [cond-mat.mes-hall]}
  \BibitemShut {NoStop}%
\bibitem [{\citenamefont {Nandkishore}(2016)}]{Nandkishore}%
  \BibitemOpen
  \bibfield  {author} {\bibinfo {author} {\bibfnamefont {R.}~\bibnamefont
  {Nandkishore}},\ }\href {\doibase 10.1103/PhysRevB.93.020506} {\bibfield
  {journal} {\bibinfo  {journal} {Phys. Rev. B}\ }\textbf {\bibinfo {volume}
  {93}},\ \bibinfo {pages} {020506} (\bibinfo {year} {2016})}\BibitemShut
  {NoStop}%
\bibitem [{\citenamefont {Sur}\ and\ \citenamefont
  {Nandkishore}(2016)}]{SurNandkishore}%
  \BibitemOpen
  \bibfield  {author} {\bibinfo {author} {\bibfnamefont {S.}~\bibnamefont
  {Sur}}\ and\ \bibinfo {author} {\bibfnamefont {R.}~\bibnamefont
  {Nandkishore}},\ }\href {http://stacks.iop.org/1367-2630/18/i=11/a=115006}
  {\bibfield  {journal} {\bibinfo  {journal} {New J. Phys.}\ }\textbf {\bibinfo
  {volume} {18}},\ \bibinfo {pages} {115006} (\bibinfo {year}
  {2016})}\BibitemShut {NoStop}%
\bibitem [{\citenamefont {Wang}\ and\ \citenamefont
  {Nandkishore}(2017)}]{WangNandkishore}%
  \BibitemOpen
  \bibfield  {author} {\bibinfo {author} {\bibfnamefont {Y.}~\bibnamefont
  {Wang}}\ and\ \bibinfo {author} {\bibfnamefont {R.~M.}\ \bibnamefont
  {Nandkishore}},\ }\href {\doibase 10.1103/PhysRevB.95.060506} {\bibfield
  {journal} {\bibinfo  {journal} {Phys. Rev. B}\ }\textbf {\bibinfo {volume}
  {95}},\ \bibinfo {pages} {060506} (\bibinfo {year} {2017})}\BibitemShut
  {NoStop}%
\bibitem [{\citenamefont {Shapourian}\ \emph {et~al.}(2018)\citenamefont
  {Shapourian}, \citenamefont {Wang},\ and\ \citenamefont {Ryu}}]{Wangetal}%
  \BibitemOpen
  \bibfield  {author} {\bibinfo {author} {\bibfnamefont {H.}~\bibnamefont
  {Shapourian}}, \bibinfo {author} {\bibfnamefont {Y.}~\bibnamefont {Wang}}, \
  and\ \bibinfo {author} {\bibfnamefont {S.}~\bibnamefont {Ryu}},\ }\href
  {\doibase 10.1103/PhysRevB.97.094508} {\bibfield  {journal} {\bibinfo
  {journal} {Phys. Rev. B}\ }\textbf {\bibinfo {volume} {97}},\ \bibinfo
  {pages} {094508} (\bibinfo {year} {2018})}\BibitemShut {NoStop}%
\bibitem [{\citenamefont {Heikkil{\"a}}\ \emph {et~al.}(2011)\citenamefont
  {Heikkil{\"a}}, \citenamefont {Kopnin},\ and\ \citenamefont
  {Volovik}}]{volovik}%
  \BibitemOpen
  \bibfield  {author} {\bibinfo {author} {\bibfnamefont {T.~T.}\ \bibnamefont
  {Heikkil{\"a}}}, \bibinfo {author} {\bibfnamefont {N.~B.}\ \bibnamefont
  {Kopnin}}, \ and\ \bibinfo {author} {\bibfnamefont {G.~E.}\ \bibnamefont
  {Volovik}},\ }\href {\doibase 10.1134/S0021364011150045} {\bibfield
  {journal} {\bibinfo  {journal} {JETP Letters}\ }\textbf {\bibinfo {volume}
  {94}},\ \bibinfo {pages} {233} (\bibinfo {year} {2011})}\BibitemShut
  {NoStop}%
\bibitem [{\citenamefont {Uchoa}\ and\ \citenamefont
  {Barlas}(2013)}]{Uchoa2013PRL}%
  \BibitemOpen
  \bibfield  {author} {\bibinfo {author} {\bibfnamefont {B.}~\bibnamefont
  {Uchoa}}\ and\ \bibinfo {author} {\bibfnamefont {Y.}~\bibnamefont {Barlas}},\
  }\href {\doibase 10.1103/PhysRevLett.111.046604} {\bibfield  {journal}
  {\bibinfo  {journal} {Phys. Rev. Lett.}\ }\textbf {\bibinfo {volume} {111}},\
  \bibinfo {pages} {046604} (\bibinfo {year} {2013})}\BibitemShut {NoStop}%
\bibitem [{\citenamefont {Altland}\ and\ \citenamefont
  {Simons}(2010)}]{AltlandSimons}%
  \BibitemOpen
  \bibfield  {author} {\bibinfo {author} {\bibfnamefont {A.}~\bibnamefont
  {Altland}}\ and\ \bibinfo {author} {\bibfnamefont {B.}~\bibnamefont
  {Simons}},\ }\href {https://books.google.com/books?id=GpF0Pgo8CqAC} {\emph
  {\bibinfo {title} {Condensed Matter Field Theory}}}\ (\bibinfo  {publisher}
  {Cambridge University Press},\ \bibinfo {year} {2010})\BibitemShut {NoStop}%
\bibitem [{\citenamefont {Coleman}(2015)}]{Coleman}%
  \BibitemOpen
  \bibfield  {author} {\bibinfo {author} {\bibfnamefont {P.}~\bibnamefont
  {Coleman}},\ }\href {https://books.google.com/books?id=PDg1HQAACAAJ} {\emph
  {\bibinfo {title} {Introduction to Many-Body Physics}}}\ (\bibinfo
  {publisher} {Cambridge University Press},\ \bibinfo {year}
  {2015})\BibitemShut {NoStop}%
\bibitem [{\citenamefont {Fulde}\ and\ \citenamefont
  {Ferrell}(1964)}]{Fulde1964PR}%
  \BibitemOpen
  \bibfield  {author} {\bibinfo {author} {\bibfnamefont {P.}~\bibnamefont
  {Fulde}}\ and\ \bibinfo {author} {\bibfnamefont {R.~A.}\ \bibnamefont
  {Ferrell}},\ }\href {\doibase 10.1103/PhysRev.135.A550} {\bibfield  {journal}
  {\bibinfo  {journal} {Phys. Rev.}\ }\textbf {\bibinfo {volume} {135}},\
  \bibinfo {pages} {A550} (\bibinfo {year} {1964})}\BibitemShut {NoStop}%
\bibitem [{\citenamefont {larkin}\ and\ \citenamefont
  {Ovchinnikov}(1964)}]{larkin1964}%
  \BibitemOpen
  \bibfield  {author} {\bibinfo {author} {\bibfnamefont {A.~I.}\ \bibnamefont
  {larkin}}\ and\ \bibinfo {author} {\bibfnamefont {Y.~N.}\ \bibnamefont
  {Ovchinnikov}},\ }\href@noop {} {\bibfield  {journal} {\bibinfo  {journal}
  {Zh. Eksp. Teor. Fiz.}\ }\textbf {\bibinfo {volume} {47}},\ \bibinfo {pages}
  {1136} (\bibinfo {year} {1964})},\ \bibinfo {note} {[Sov. Phys.
  JETP20,762(1965)]}\BibitemShut {NoStop}%
\end{thebibliography}%

\end{document}